\documentclass[
reprint,
showkeys,
amsmath,amssymb,
aps,
prr,
]{revtex4-2}

\usepackage{graphicx}
\usepackage{dcolumn}
\usepackage{bm}
\usepackage{hyperref}
\usepackage{pdfpages}
\usepackage{subfiles}
\usepackage{etoolbox} 
\makeatletter
\patchcmd{\@outputpage@head}{\@ifx{\LS@rot\@undefined}{}{\LS@rot}}{}{}{}
\makeatother

\begin{document}


\title{Neuronal correlations shape the scaling behavior of memory capacity and nonlinear computational capability of reservoir recurrent neural networks}

\author{Shotaro Takasu}
 \email{shotaro.takasu.63x@st.kyoto-u.ac.jp}
\author{Toshio Aoyagi}%
\affiliation{%
  Graduate School of Informatics, Kyoto University, Yoshida-Honmachi, Sakyo-ku, Kyoto 606-8501, Japan
}%

\date{\today}

\begin{abstract}
Reservoir computing is a powerful framework for real-time information processing, characterized by its high computational ability and quick learning, with applications ranging from machine learning to biological systems. In this paper, we investigate how the computational ability of reservoir recurrent neural networks (RNNs) scales with an increasing number of readout neurons. First, we demonstrate that the memory capacity of a reservoir RNN scales sublinearly with the number of readout neurons. To elucidate this observation, we develop a theoretical framework for analytically deriving memory capacity that incorporates the effect of neuronal correlations, which have been ignored in prior theoretical work for analytical simplicity. Our theory successfully relates the sublinear scaling of memory capacity to the strength of neuronal correlations. Furthermore, we show this principle holds across diverse types of RNNs, even those beyond the direct applicability of our theory. Next, we numerically investigate the scaling behavior of nonlinear computational ability, which, alongside memory capacity, is crucial for overall computational performance. Our numerical simulations reveal that as memory capacity growth becomes sublinear, increasing the number of readout neurons successively enables nonlinear processing at progressively higher polynomial orders. Our theoretical framework suggests that neuronal correlations govern not only memory capacity but also the sequential growth of nonlinear computational capabilities. Our findings establish a foundation for designing scalable and cost-effective reservoir computing, providing novel insights into the interplay among neuronal correlations, linear memory, and nonlinear processing.
\end{abstract}

\maketitle

\section{Introduction}
Over the past decade, deep learning has advanced significantly~\cite{Mienye2024}, with generative AI in particular beginning to transform many aspects of daily life~\cite{Sengar2025}. Although deep neural networks possess the capacity to represent highly complex functions~\cite{Cybenko1989, Raghu2017}, they typically require substantial computational resources, large volumes of training data, and long training times~\cite{Thompson2022}. These high computational demands pose a major obstacle to applying deep learning in real-time computing on edge-computing devices~\cite{Marchisio2019}.

Reservoir computing (RC) is a machine learning framework for efficiently training large-scale recurrent neural networks (RNNs), referred to as "reservoirs" in the context of RC~\cite{Jaeger2001, Maass2002}. In contrast to conventional training methods such as backpropagation through time (BPTT), RC optimizes only readout weights and leaves the remaining weights fixed, which enables quick and low-cost learning. Despite its training simplicity, RC has high computational performance and a broad range of applications for processing time-series data~\cite{Yan2024}. RC is not restricted to RNNs, and a wide variety of dynamical systems can be utilized as reservoirs under appropriate conditions. In particular, RC using a real physical system such as soft matter and optical systems, so called "physical reservoir computing", has been intensively studied in recent years~\cite{Nakajima}.

In a typical RC setting, the readout connections linking reservoir units to output units are sparse~\cite{Nakajima}. This means that the size of the reservoir ($N$) is substantially larger than the number of readout units ($L$) that connect to the output units. The performance of RC empirically improves as $L$ increases because this increases the expressive power of the reservoir output~\cite{Cucchi2022}. In contrast, implementing a large number of readout connections can be resource-intensive, especially in physical RC implementations. Consequently, understanding the relationship between $L$ and the computational ability of a reservoir is essential for constructing cost-effective RC. Despite its practical importance, however, a guiding principle for determining an appropriate $L$ has yet to be fully explored. 

In this study, we aim to address this gap by elucidating how memory capacity~\cite{Jaeger2002} and nonlinear computational ability~\cite{Dambre2012} -- two key determinants of a reservoir's computational power -- scale with $L$, with the particular focus on the role of neuronal correlations. We demonstrate, for the first time, that the memory capacity of an RNN increases monotonically as increasing $L$, with its growth rate progressively declining (Fig.\ref{fig:overview}). This {\it sublinear} scaling of memory capacity cannot be explained by the previous theoretical works assuming a scaling regime where $L$ does not scale with $N$ , i.e. $L \sim O(1)$~\cite{Schuecker2018, Haruna2019}, because the memory capacity in this regime exhibits linear scaling, as will be shown later. To bridge the gap between our observations and existing theory, we develop a novel theory for analytically deriving memory capacity for $L \sim O(\sqrt{N})$. Using our theoretical framework, we demonstrate that the correlations between reservoir neurons, despite being very weak with a magnitude scaling as $O(1/\sqrt{N})$, play an important role in the declining growth rate of memory capacity. Furthermore, numerical simulations show that the ability to perform more complex nonlinear tasks emerges supralinearly and sequentially as $L$ increases. These findings indicate that the number of readout connections should be tailored based on the specific memory and nonlinearity requirements of the task at hand.

\section{Model setup} \label{sec2}
We investigate computational capacity of RC using a large random RNN as a reservoir (Fig.\ref{fig:overview}(a)), which is known as Echo State Network (ESN)~\cite{Jaeger2001}, one of the canonical models for RC ~\cite{Lukosevicius2012, Sussilo2009}. The state of $i$-th neuron at discrete time $t$ is described by the variable $x_i(t)$. The activation function $\phi$ is assumed to be an odd saturating sigmoid function satisfying $\phi'(0)=1$, $\phi'(x)>0$, $\phi''(x) \leq 0\ (x\geq 0)$ and $\phi(\pm \infty)=\pm 1$.  The time evolution of a reservoir RNN is determined by the difference equation,
\begin{eqnarray} 
\label{model}
x_i(t) = \sum_{j=1}^N J_{ij}\phi(x_j(t-1)) + u_i s(t) + \xi_i(t),
\end{eqnarray}
where $N\gg 1$ indicates the total number of neurons. The recurrent weights, $J_{ij}$, and the input weights, $u_i$, are sampled i.i.d from Gaussian distributions, i.e. $J_{ij} \sim \mathcal{N}(0, g^2/N)$ and $u_i \sim \mathcal{N}(0,1)$. The parameter $g$ controls the strength of the recurrent weights. The input signal at time $t$, represented by $s(t)$, is Gaussian white noise with zero mean and variance $\sigma_s^2$. To take into account inherent neuronal noise, each neuron is subjected to independent Gaussian white noise, $\xi_i(t)$, with zero mean and variance $\sigma_n^2$. Notably, the input signal -- shared across all neurons -- induces correlations among reservoir neurons, whereas the independent neuronal noise acts to diminish them.

The macroscopic dynamical behavior of the large random RNN described by Eq.(\ref{model}) has been extensively studied by means of dynamical mean-field theory (DMFT)~\cite{Sompolinsky1988}. In the absence of the inputs and noise, i.e. $\sigma_s^2=\sigma_n^2=0$, the RNN exhibits phase transition from zero fixed-point to chaotic dynamics at $g=1$ in the large network size limit $N\to \infty$~\cite{Sompolinsky1988, Molgedey1992}. In the presence of inputs ($\sigma_s^2>0$ or $\sigma_n^2>0$), chaotic variability is typically suppressed~\cite{Schuecker2018,Molgedey1992,Engelken2022,Takasu2024,Haruna2019, Massar2013}, while it has recently been shown that in RNNs incorporating gating mechanisms similar to those in gated recurrent units (GRUs), external inputs can instead induce chaos~\cite{Krishnamurthy2022}. 

The output of the reservoir RNN is defined as $z(t)=\sum_{i=1}^L w_i x_i(t)$, where $L$ is the number of readout neurons. We assume that the readout neurons are sparse, i.e., $L\ll N$. According to the RC framework, the output weights, $\{w_i\}_{i=1}^L$, are optimized to minimize the time-averaged squared error between the output signals, $\{z(t)\}_{t=1}^T$, and the desired signals, $\{f(t)\}_{t=1}^T$. Since this is a least-squares problem, the solution for $\bm{w} \in \mathbb{R}^L$ is explicitly given by
\begin{eqnarray} 
\label{solution for w_out}
\bm{w} = (XX^\top)^{-1} X \bm{f},
\end{eqnarray}
where the matrix $X \in \mathbb{R}^{L \times T}$ contains the history of readout neuron activities throughout simulation time length $T$, and the vector $\bm{f} \in \mathbb{R}^{T}$ denotes the target time-series.

\section{Memory capacity}

{\it Memory capacity}~\cite{Jaeger2002,Dambre2012} is a commonly used benchmark for RC, quantifying how accurately a reservoir can reproduce its past Gaussian white noise inputs. Specifically, it is defined as $MC \equiv \sum_{d=0}^\infty M_d$, where
\begin{eqnarray} \label{Md}
M_d \equiv 1-\frac{{\rm{min}}_{\bm{w}}\langle (z(t)-s(t-d))^2 \rangle}{\langle s(t)^2 \rangle},
\end{eqnarray}
and the angular bracket denotes time averaging. Here, the optimized readout weights $\bm{w}$ are obtained by substituting $f(t)=s(t-d)$ into Eq.(\ref{solution for w_out}). A high value of $M_d$ indicates that the reservoir can accurately output its input signal from $d$ steps prior. It is shown that $0\leq M_d \leq 1$ holds true~\cite{Jaeger2002,Dambre2012}. Memory capacity is defined as the infinite sum of $M_d$, but it has been proven to be finite and bounded by the number of readout neurons, i.e. $0\leq MC \leq L$ ~\cite{Jaeger2002, Dambre2012}. In particular, for RC with a linear activation function operating in noise-free conditions ($\sigma_n^2=0$), memory capacity is equivalent to $L$, independently of $g$ and $\sigma_s^2$ ~\cite{Jaeger2002}.

\begin{figure}[tbp]
    \centering
    \includegraphics[width=0.8\linewidth]{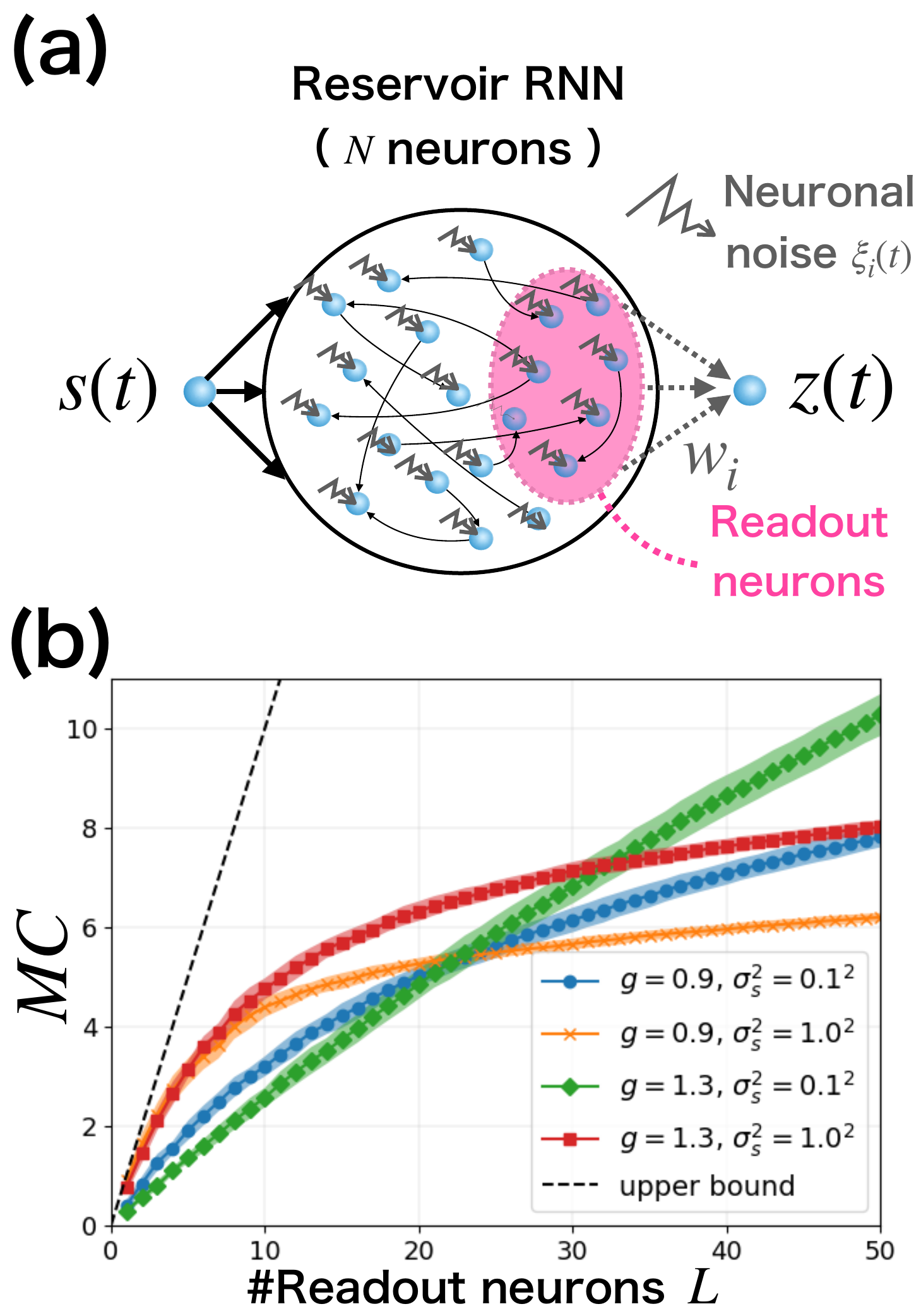}
    \caption{(a) Overview of RC with a random RNN receiving input signals and inherent neuronal noise. The shaded region represents the readout neurons. (b) Numerical simulations for memory capacity ($MC$) of the reservoir RNNs with network size $N=10000$, noise intensity $\sigma_n^2 = 0.1^2$, and activation function $\phi(x) = \tanh(x)$. Shaded area represents mean$\pm$std of direct numerical simulations for 10 different network and input signal realizations. The dashed line indicates the theoretical upper bound of memory capacity, i.e., $MC=L$. Each $MC$ curve shows a sublinear scaling with increasing $L$. In the simulations, the sum of $M_d$ is calculated up to $d=10^3$. Simulation time length is $T=10^4$, and a washout period of $T_{\rm washout} = 10^3$ steps is used to discard initial transients. 
    }
    \label{fig:overview}
\end{figure}

Fig.\ref{fig:overview}(b) illustrates the relationship between $L$ and the memory capacity for the reservoir RNN described by Eq.(\ref{model}), obtained through numerical simulations. As shown, memory capacity grows monotonically, but its growth rate gradually diminishes as $L$ increases. Consequently, memory capacity can be characterized as a sublinear function of $L$. 

\section{Theoretical analysis of memory capacity}

To elucidate the mechanism behind the decaying growth rate of memory capacity, we provide a novel theory for deriving an analytical solution of memory capacity. Our approach incorporates some concepts and methods from statistical physics. Hereafter, we make an additional assumption regarding the scale of the model parameters:
\begin{eqnarray} \label{scaling}
L=\alpha \sqrt{N},\ \  
\sigma_s^2 = \frac{\tilde{\sigma}_s^2}{\sqrt{N}},\ \ 
\sigma_n^2 \sim O(1),
\end{eqnarray}
where both $\alpha$ and $\tilde{\sigma}_s^2$ are of $O(1)$. The parameter $\alpha$ represents the relative magnitude of $L$ compared to $\sqrt{N}$. In contrast to previous studies~\cite{Toyoizumi2011,Schuecker2018,Haruna2019}, where $L\sim O(1)$, our model employs a significantly larger number of readout neurons, although the output connections remain sparse, as $L=\alpha \sqrt{N}\ll N$. It should be noted that while our theoretical analysis in this study is conducted under this assumption, our theoretical framework can be extended to more general case where $L\sim O(N^c)$ and $\sigma_s^2 \sim O(1/N^c)$ for $0 \leq c \leq 1/2$ (see Appendix \ref{appendixB2}).

Substituting the optimal solution for $\bm{w}$ from Eq.(\ref{solution for w_out}) into the definition of $M_d$ (Eq.(\ref{Md})) yields:
\begin{eqnarray} \label{Md2}
M_d = \frac{\bm{a}_d^\top C^{-1} \bm{a}_d}{\langle s(t)^2 \rangle},
\end{eqnarray}
where the elements of $\bm{a}_d\in \mathbb{R}^L$ and $C \in \mathbb{R}^{L\times L}$ are respectively $(\bm{a}_d)_i\equiv \langle s(t-d) x_i(t)\rangle$ and $C_{ij} \equiv \langle x_i(t) x_j(t)\rangle$ ($i,j$ indicate the indices of the readout neurons)~\cite{Dambre2012}. The inverse of the covariance matrix, $C^{-1}$, prevents us from making progress on analytical calculation. Previous theoretical studies circumvented this issue by assuming the off-diagonal entries of $C$ are zero, that is, ignoring neuronal correlations~\cite{Toyoizumi2011, Schuecker2018, Haruna2019}. However, as discussed later, this approximation results in linear scaling of memory capacity, $MC \propto L$, conflicting with our numerical simulations as shown in Fig.\ref{fig:overview}(b).

We notice that from the scale assumption Eq.(\ref{scaling}), the diagonal entries of $C$ are of $O(1)$, whereas the non-diagonal ones are of $O(1/\sqrt{N})$ (See Appendix~\ref{appendixA2}). Hence, the non-diagonal elements are much smaller than the diagonal ones, which motivates us to perform Neumann series expansion,
\begin{eqnarray} \label{neumann}
C^{-1} = (C_{\rm diag} + C_{\rm nondiag})^{-1} 
= \sum_{n=0}^\infty C_{\rm diag}^{-1}\left( - C_{\rm nondiag} C_{\rm diag}^{-1} \right)^n, \nonumber\\
\end{eqnarray}
where $(C_{\rm diag})_{ij}\equiv \delta_{ij} \langle x_i(t)^2 \rangle$ and $(C_{\rm nondiag})_{ij}\equiv (1-\delta_{ij})\langle x_i(t) x_j(t) \rangle$. This approach allows us to circumvent the issue of the inverse matrix and enables an analytical calculation of $M_d$.

In the large network size limit, we can assume self-averaging for $M_d$, i.e., $\lim_{N\to \infty} M_d = \lim_{N\to\infty}[M_d] $, where the square bracket denotes the average over network realizations, known as {\it quenched average}~\cite{Helias}. Consequently, substituting Eq.(\ref{neumann}) into Eq.(\ref{Md2}) and taking quenched average of $M_d$, we analytically obtain memory capacity as:
\begin{eqnarray} \label{mc theory}
&&\lim_{N\to \infty} MC(L=\alpha\sqrt{N}) =\lim_{N\to\infty}\sum_{d=0}^\infty \left[M_d(L=\alpha\sqrt{N})\right]\\
&&= \sum_{d=0}^\infty \sum_{n=0}^\infty
(-1)^{n} \left\{
\frac{\alpha \tilde{\sigma}_s^2}{[\langle x_i^2\rangle]}  
\left(
g\langle \phi'(x) \rangle_{x\sim \mathcal{N}(0,[\langle x_i^2\rangle])}
\right)^{2d}
\right\}^{n+1} \nonumber
\end{eqnarray}
where the value of $[\langle x_i^2\rangle]$ can be obtained by solving a self-consistent equation (dynamical mean-field equation~\cite{Sompolinsky1988}),
\begin{eqnarray}\label{DMFeq}
[\langle x_i^2\rangle] = \sigma_n^2 + g^2 \langle \phi(x)^2 \rangle_{x\sim \mathcal{N}(0,[\langle x_i^2\rangle])}.
\end{eqnarray}
The memory capacity of a linear RNN with neuronal noise can be derived in a similar fashion, and this derivation also reveals a sublinear scaling (see Appendix \ref{appendixB4}).

The derivation of these results requires calculating quenched averages of terms involving neuronal correlations, such as 
\begin{eqnarray}
    \left[\langle s(t-d)x_i(t)\rangle \langle x_i(t)x_j(t) \rangle \langle s(t-d)x_j(t)\rangle \right]\ \ (i\neq j).
\end{eqnarray}
However, these quantities cannot be evaluated using DMFT because it reduces the original $N$-body system to an effective single-body description, thereby neglecting inter-neuronal correlations. To overcome this limitation, we adopt the dynamical cavity method, a technique originally introduced by Clark et al.~\cite{Clark2022} from statistical physics to the analysis of neural networks. A detailed description of this theory is provided in Appendix~\ref{appendixA1} and \ref{appendixA2}, followed by the full derivation of Eq.(\ref{mc theory}) in Appendix~\ref{appendixB1}. In the following, we use the shorthand notation, $\langle f(x)\rangle_* \equiv \langle f(x) \rangle_{x\sim \mathcal{N}(0,[\langle x_i^2\rangle])}$. 

When the activation function is an error function, $\phi(x) = \int_0^x e^{-\frac{\pi}{4} t^2}dt$,  we can analytically integrate both $\langle \phi'(x) \rangle_*$ and $\langle \phi(x)^2 \rangle_*$, yielding 
\begin{eqnarray}\label{mc theory for erf}
MC = \sum_{d=0}^\infty \sum_{n=0}^\infty
(-1)^{n}
\left\{
\frac{\alpha \tilde{\sigma}_s^2}{[\langle x_i^2\rangle]}  
\left( \frac{g^2}{1+\frac{\pi}{2}[\langle x_i^2\rangle]}
\right)^{d}
\right\}^{n+1},\nonumber \\
\end{eqnarray}
where $[\langle x_i^2\rangle]$ is determined by solving
\begin{eqnarray}\label{DMFeq for erf}
[\langle x_i^2\rangle] = \sigma_n^2 + g^2 \left(-1 + \frac{4}{\pi}\arctan\sqrt{1+\pi [\langle x_i^2\rangle]} \right). \nonumber \\
\end{eqnarray}

\begin{figure}[tbp]
    \centering
    \includegraphics[width=\linewidth]{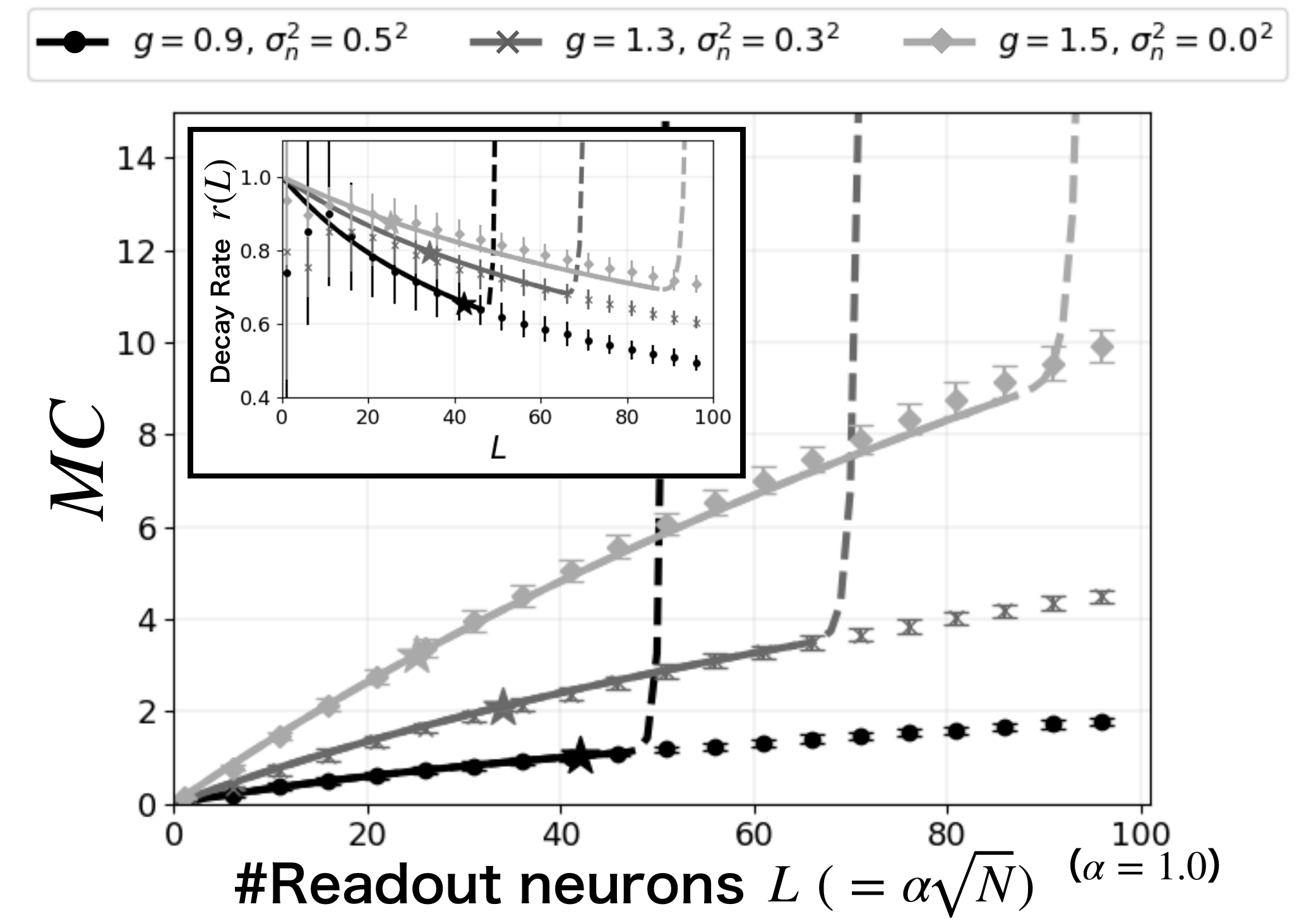}
    \caption{Memory capacity as a function of $L$. An inset shows corresponding decay rate $r(L)$. Solid curves represent the analytical solutions given by Eqs.(\ref{mc theory for erf})(\ref{DMFeq for erf}), while error bars indicate mean$\pm$std for 10 networks and inputs realizations obtained through numerical simulations. Star marks represent the upper bounds of the sufficient conditions for the convergence of the series expansion. Dashed lines indicate divergence of theoretical values.  The theoretical values and simulation results show excellent agreement in the region where the analytical solution converges. The activation function is an error function, $\phi(x) = \int_0^x e^{-\frac{\pi}{4} t^2}dt$. The input intensity is $\sigma_s^2 = 1.0^2 / \sqrt{N}$ ($\tilde{\sigma}_s^2 = 1.0^2$). The network size is $N=10000$, and simulation time length is $T=10^4$. The transient washout time is $T_{\rm washout}=10^3$.
    }
    \label{fig:analytical MC}
\end{figure}

We confirm that the analytical values of $MC$ are consistent with the numerical simulations, exhibiting a sublinear function of $L$, for $L$ below a threshold, as illustrated in Fig.\ref{fig:analytical MC}. When $L$ exceeds the threshold, the analytical values rapidly deviate from the simulated ones and eventually diverge (dashed lines in Fig.\ref{fig:analytical MC}). This divergence occurs because the Neumann series expansion (Eq.(\ref{neumann})) fails to converge when the spectral norm $\| C_{\rm nondiag} C_{\rm diag}^{-1} \|$ exceeds one for large $L$. Evaluating this spectral norm to determine the convergence threshold for $L$ is a challenging task. Instead, we note that the spectral norm is bounded by the Frobenius norm $\| \cdot \|_F$ and that $\| C_{\rm nondiag} C_{\rm diag}^{-1} \|_F$ can be obtained analytically. We therefore derive a sufficient condition for the series expansion to converge (see Appendix~\ref{appendixB1}). The corresponding upper bounds of $L$ satisfying this condition are indicated by star marks in Fig.\ref{fig:analytical MC}.

\section{Neuronal correlations shape the sublinear scaling of memory capacity}

Based on our analytical solution for memory capacity, we can now investigate the underlying reason for its sublinear scaling with $L$. To quantify the extent to which the growth rate of memory capacity decays, we define a {\it decay rate}, 
\begin{eqnarray} \label{decay rate}
r(L) \equiv \frac{MC(L)}{L \times MC(1)}. 
\end{eqnarray}
By this definition, if $MC(L)$ scales linearly with $L$, then $r(L)=1$ holds for all $L$. In contrast, $r(L)$ decays from unity when $MC(L)$ is a sublinear function of $L$. As depicted in the inset of Fig.\ref{fig:analytical MC}, $r(L)$ consistently decays for various parameter combinations, confirming the sublinear scaling of memory capacity with respect to $L$. 
 
\begin{figure}[htbp]
    \centering
    \includegraphics[width=0.8\linewidth]{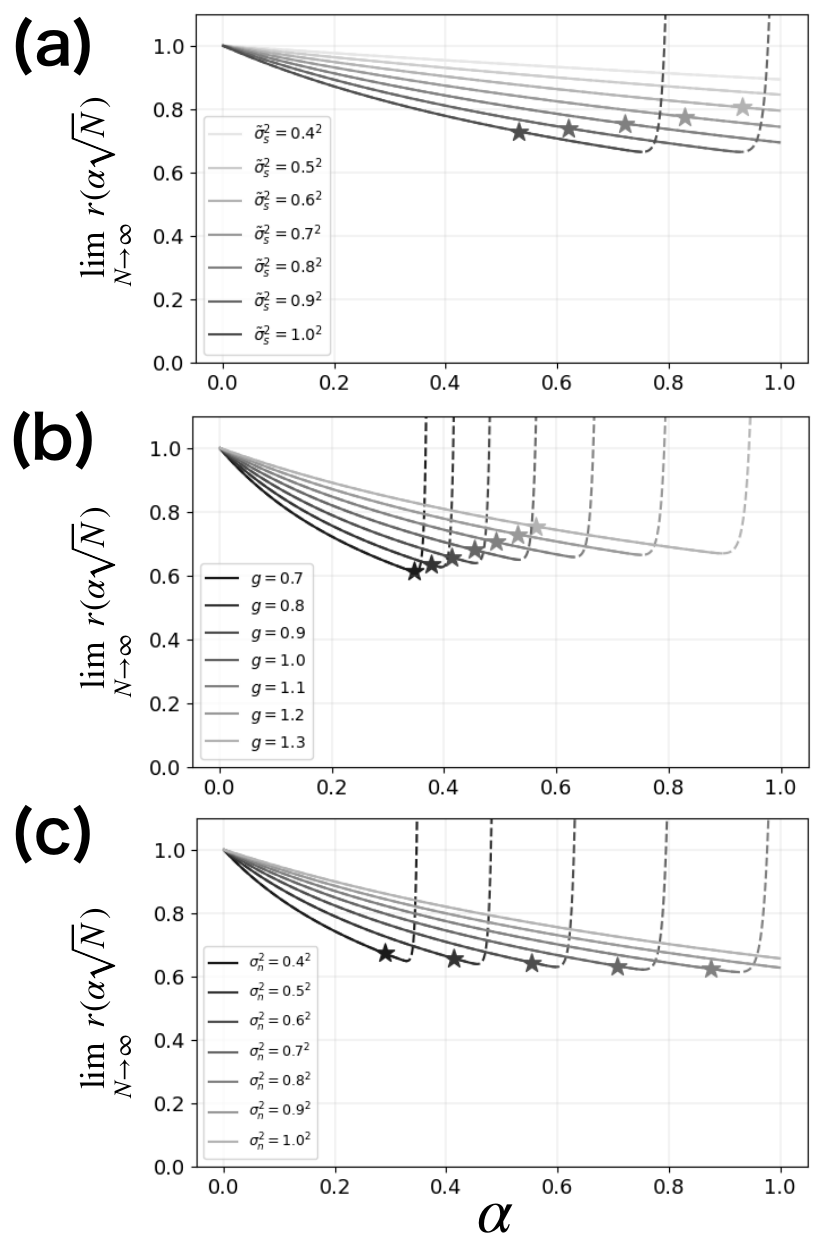}
    \caption{Analytical solutions for the decay rate of memory capacity as a function of $\alpha$. The activation function is an error function, $\phi(x) = \int_0^x e^{-\frac{\pi}{4} t^2}dt$. Each figure varies a single parameter: (a) input intensity $\tilde{\sigma}_s^2$, (b) recurrent weight scale $g$, (c) noise intensity $\sigma_n^2$, while the remaining parameters are fixed ($g=1.2$, $\tilde{\sigma}_s^2 = 1.0^2$, $\sigma_n^2=0.5^2$). A gradient from darker to lighter gray lines indicates a decreasing level of neuronal correlations. The decay rate for systems with strong neuronal correlations decreases more rapidly.}
    \label{fig:decaying rate bahavior}
\end{figure}

From our analytical solution for memory capacity (Eq.(\ref{mc theory})), we obtain the analytical form for the decay rate (see Appendix~\ref{appendixB3} for the derivation) as:
\begin{eqnarray} \label{analytical r}
\lim_{N\to \infty} r(L=\alpha\sqrt{N}) &&= 1 - \sum_{n=1}^\infty (-1)^{n-1}\left( \frac{\tilde{\sigma}_s^2}{[\langle x_i^2 \rangle]} \alpha \right)^{n}  \\
&&\hspace{60pt}\times\frac{1- (g\langle \phi'(x) \rangle_*)^2}{1- (g\langle \phi'(x) \rangle_*)^{2n+2}}. \nonumber
\end{eqnarray}
Notably, the $n$-th term corresponds exactly to the $n$-th term in the Neumann series expansion, Eq.(\ref{neumann}). Therefore, under the assumption of vanishing neuronal correlations ($C_{\rm nondiag}=O$), the summation terms in Eq.(\ref{analytical r}) disappear, resulting in $r(L)=1$, and thus, linear scaling. This implies that higher neuronal correlations contribute to the faster decay of $r(L)$ via the summation terms in Eq.(\ref{analytical r}). In fact, we confirm that $r(L)$ decays more rapidly for hyperparameters that lead to higher neuronal correlation, such as larger input signals ($\tilde{\sigma}_s^2$), smaller recurrent weights ($g$), and lower noise intensity ($\sigma_n^2$), as shown in Fig.\ref{fig:decaying rate bahavior}. 

Importantly, while the slope of the memory capacity curve at the origin, which is proportional to $MC(1)$, reaches its maximum when the reservoir RNN operates near the edge of chaos~\cite{Schuecker2018,Haruna2019}, the extent of the decay of $r(L)$ varies monotonically with the hyperparameters $g$, $\sigma_s^2$ and $\sigma_n^2$, irrespective of whether the RNN is near the edge of chaos. This finding suggests a new perspective: instead of proximity to the edge of chaos, neuronal correlation is a novel mechanism that governs the computational ability of reservoir computing.

\begin{figure}[t]
    \centering
    \includegraphics[width=0.8\linewidth]{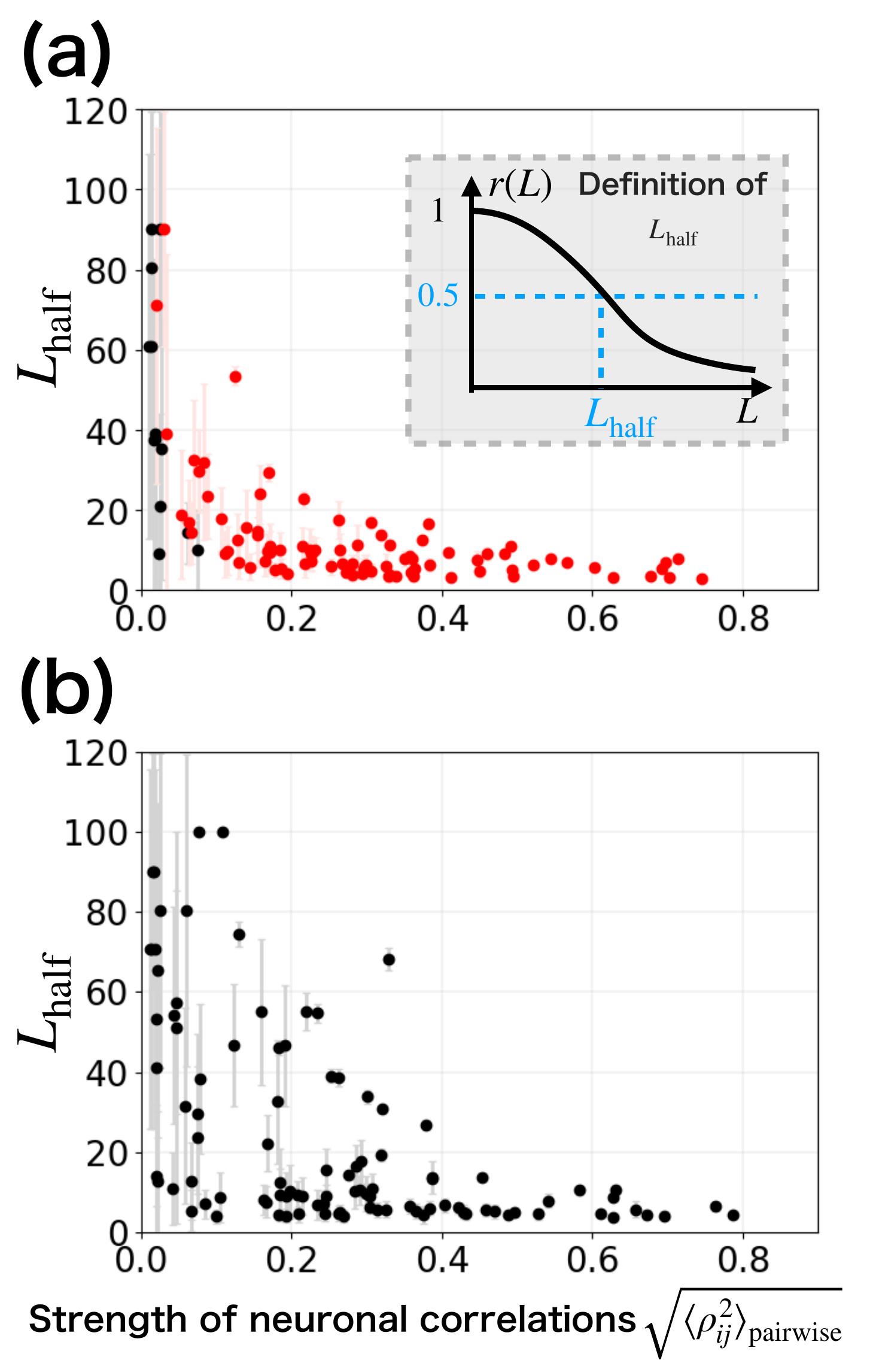}
    \caption{ Relationship between neuronal correlation and the half-life of memory capacity growth. The x-axis represents the level of neuronal correlations quantified by the root mean square (RMS) of pairwise Pearson correlation coefficients. The y-axis denotes $L_{\rm half}$, the definition of which is graphically shown in the inset of panel (a). (a) Results for RNNs with $\phi=\tanh$. Each data point corresponds to a set of hyperparameters sampled from $g\sim U(0.2, 3)$, $\sigma_s \sim U(0.1, 3)$ and $\sigma_n \sim U(0, 3)$. Red points indicate parameter regimes beyond our theoretical framework, as determined by $L_{\rm half}$ surpassing the deviation threshold of the theoretical values of $MC$ (Fig.\ref{fig:analytical MC}). (b) Results for RNNs with $\phi(x)= \max (0, x)$ (ReLU). Each data point corresponds to a set of hyperparameters sampled from $g\sim U(0.2, \sqrt{2})$, $\sigma_s \sim U(0.1, 3)$ and $\sigma_n \sim U(0, 3)$. For both (a) and (b), a total of 100 samples are plotted. Each data point represents the mean value over 10 network and input realizations. Error bars indicate the standard deviation for both $L_{\rm half}$ and $\sqrt{\langle \rho_{ij}^2 \rangle_{\rm pairwise}}$, but most of the error bars for the latter are too small to be visible behind the data points. Overall, for both activation functions, $L_{\rm half}$ tends to be smaller for stronger neuronal correlations. All simulations use a network size of $N=10000$ and a simulation time of $T=10^4$. The transient washout time is $T_{\rm washout}=10^3$.
    }
    \label{fig:rho_Lhalf}
\end{figure}

We have thus far derived the memory capacity under the specific set of scaling assumptions (Eq.(\ref{scaling})). However, practical applications often deviate from these conditions, raising the question of whether neuronal correlations remain a key determinant of sublinear memory capacity growth when these assumptions are relaxed. To explore this, Fig.\ref{fig:rho_Lhalf}(a) examines a broader range of model parameters, including larger input intensities, various noise levels, and different values of $g$. Here, we introduced a {\it half-life} of memory capacity growth, $L_{\rm half}$, defined as the value of $L$ at which $r(L)$ equals $0.5$ (Fig.\ref{fig:rho_Lhalf}(a), inset). A lower $L_{\rm half}$ indicates more rapid decay of $r(L)$. The overall strength of neuronal correlations is quantified by the root mean square (RMS) of pairwise Pearson correlation coefficients, $\sqrt{\langle \rho_{ij}^2 \rangle_{\rm pairwise}}$, where 
\begin{eqnarray}
    \rho_{ij} \equiv \frac{\langle x_i x_j \rangle - \langle x_i\rangle \langle x_j\rangle}{\sqrt{\left(\langle x_i^2 \rangle- \langle x_i \rangle^2\right)\left(\langle x_j^2 \rangle- \langle x_j \rangle^2\right)}}. \nonumber
\end{eqnarray}
Note that due to the symmetry of the weight distribution, the time-averaged value $\langle x_i \rangle$ almost vanishes for all neurons. We use the RMS value because the symmetry of the weight distribution creates a pairwise distribution of $\rho_{ij}$ values that is symmetric around zero. Consequently, the simple average $\langle \rho_{ij} \rangle_{\rm pairwise}$ is always zero and thus uninformative, whereas the RMS value properly quantifies the typical magnitude of the neuronal correlations.

As shown in Fig.\ref{fig:rho_Lhalf}(a), strong neuronal correlations are indeed closely associated with a more rapid decay of memory capacity growth, with $L_{\rm half}$ being inversely proportional to the level of neuronal correlations. This finding highlights the persistent influence of neuronal correlations even when our original scaling assumptions are relaxed.

\section{Robustness of our findings across other types of RNNs}

Thus far, we have examined the scaling behavior of reservoir RNNs with a sigmoid activation function and recurrent weights drawn i.i.d from a Gaussian distribution. We found that the level of neuronal correlations, which is controlled by hyperparameters such as $g$, $\sigma_s^2$, and $\sigma_n^2$ in our model, determines the extent to which the growth rate of memory capacity decays. While our model is widely used in both theoretical~\cite{Schuecker2018, Haruna2019, Inubushi2017} and practical~\cite{Lukosevicius2012} settings, a natural question arises: can these findings be generalized to other types of RNNs? Here, we hypothesize that our findings on the relationship between neuronal correlations and memory capacity hold more generally. To test this hypothesis, we numerically study three types of RNNs: those with a ReLU activation function, a heavy-tailed distribution of recurrent weights, and reciprocal motifs.

\subsection{RNN with a ReLU activation function} \label{sec:relu}
We first study the neuronal correlations and memory capacity of an RNN described by Eq.(\ref{model}) with $\phi(x) = \max(0,x)$ (ReLU). All parameter settings are the same as those described in Section \ref{sec2}, with the exception that $g$ must be lower than $g_c = \sqrt{2}$ because the system diverges for $g>g_c$. To demonstrate this, we first derive the dynamical mean-field equation for this system in the limit of $N\to\infty$. The derivation is the same as in Appendix \ref{appendixA1} and yields the following equation:
\begin{eqnarray}
[x_i(t+1)^2] = \frac{g^2}{2}[x_i(t)^2] + s(t+1)^2 + \sigma_n^2.
\end{eqnarray}
This self-consistent equation admits a stable solution of $[x_i(t)^2]$ only when $g<g_c=\sqrt{2}$. For $g>g_c$, the system diverges regardless of the values of $s(t)$ and $\sigma_n$.

Our theoretical framework for the analytical derivation of neuronal correlations and memory capacity cannot be applied to an RNN with the ReLU activation function. This limitation arises because our theory utilizes Price's theorem (Eq.(\ref{Price's theorem})), which cannot be applied to the ReLU function as it is non-smooth at the origin. Therefore, we numerically calculate neuronal correlations and memory capacity, and plot them to examine their relationship. 

Following the same procedure as in the previous section, we numerically calculated the strength of neuronal correlations and $L_{\rm half}$. The results, shown in Fig. \ref{fig:rho_Lhalf}(b), reveal a negative correlation between the level of neuronal correlations and $L_{\rm half}$. Although this correlation is less pronounced than for the sigmoid network (Fig. \ref{fig:rho_Lhalf}(a)), the finding nonetheless supports our hypothesis that neuronal correlations govern the scaling behavior of memory capacity.

\subsection{RNN with a heavy-tailed distribution of recurrent weights} \label{sec:cauchy}

We now turn to the case of an RNN where the distribution of recurrent weights is heavy-tailed or scale-free.  Recent experimental work has revealed that the distribution of synaptic strength in the brain follows a heavy-tailed distribution across a wide range of brain areas~\cite{Mizuseki2014}. Several theoretical studies have also proposed potential roles for heavy-tailed synapse distributions in brain function~\cite{Teramae2012,Kusmierz2020}.

The theoretical analysis of dynamics in large random RNNs, using methods like dynamical mean-field theory or the dynamical cavity method, typically relies on the central limit theorem. This allows the statistical behavior of neurons to be described as a Gaussian process (see Appendix~\ref{appendixA}). However, this assumption breaks down for heavy-tailed weight distributions, where the central limit theorem is not guaranteed to hold. This leads to a fundamental difference in dynamical properties between conventional RNNs and those with heavy-tailed synapses. Although a theoretical framework to derive the dynamical mean-field equations for such RNNs has recently been proposed using the generalized central limit theorem~\cite{Wardak2022}, the analysis of neuronal correlations remains challenging.  Consequently, we must rely on numerical simulations to investigate the neuronal correlations and memory capacity of these RNNs.

The model examined here is identical to that described in Section~\ref{sec2}, with the exception that the recurrent weights are i.i.d drawn from a Cauchy distribution~\cite{Kusmierz2020}:
\begin{eqnarray} \label{Cauchy dist}
    p(J_{ij}) = \frac{1}{\pi} \frac{\gamma/N}{(\gamma/N)^2 + J_{ij}^2}.
\end{eqnarray}
Here, the hyperparameter $\gamma>0$ determines the strength of recurrent weights. In the limit of $N\to \infty$, the eigenvalue spectrum of $J$ is unbounded and its spectral radius is infinite for any $\gamma>0$~\cite{Wardak2022}, leading to chaotic dynamics in the RNN regardless of $\gamma$. On the other hand, when $N$ is finite, whether the dynamics are chaotic or stable depends on the specific realization of $J$~\cite{Kusmierz2025}. 

\begin{figure}[htbp]
    \centering
    \includegraphics[width=0.9\linewidth]{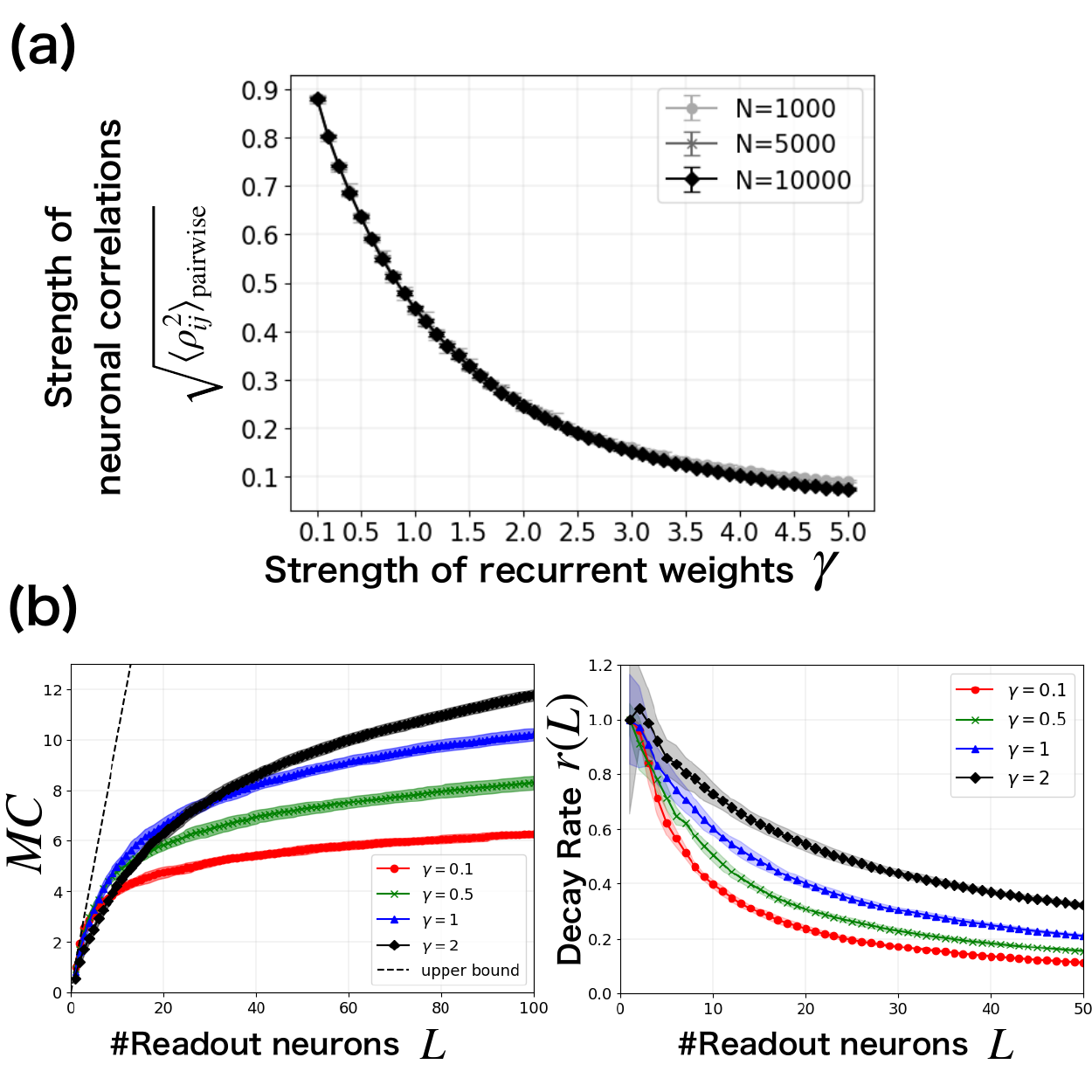}
    \caption{Numerical simulations of neuronal correlations and memory capacity of RNNs with recurrent weights drawn from a Cauchy distribution. (a) The level of neuronal correlations, quantified by the root mean square of pairwise Pearson correlation coefficients, as a function of the scale parameter $\gamma$. Data are shown for network sizes $N=1000,5000,$ and $10000$. (b) Left: Memory capacity (MC) as a function of the number of readout neurons, $L$. Right: Decay rate ($r(L)$) as a function of $L$. Both plots show data for different values of $\gamma$ with a fixed network size of $N=10000$. Networks with a large $\gamma$ exhibit weaker neuronal correlations and, consequently, a slower decay of $r(L)$.
    For both (a) and (b), the simulation time length is $T=10^4$. The input and noise intensity are $\sigma_s^2=1.0$ and $\sigma_n^2 = 0.0$, respectively. The activation function is $\phi=\tanh$. Errorbars and shaded areas represent mean$\pm$std over 10 network and input realizations. }
    \label{fig:cauchy}
\end{figure}

As demonstrated in Fig.\ref{fig:cauchy}(a), the level of neuronal correlations is controlled by the parameter $\gamma$, analogous to the role of the parameter $g$ in our canonical Gaussian model. These neuronal correlations are considered to be of $O(1)$ with respect to the network size $N$, as indicated by the nearly constant values of $\sqrt{\langle \rho_{ij}^2} \rangle_{\rm pairwise}$ for increasing $N$.

Since we have observed that larger $\gamma$ is associated with weaker neuronal correlations, our hypothesis predicts that the growth rate of memory capacity decays more slowly for larger $\gamma$. This prediction is supported by the results in Fig.~\ref{fig:cauchy}(b), which shows that the decay rate $r(L)$ decreases more slowly as $\gamma$ increases. Therefore, we conclude that our findings generalize to RNNs with a heavy-tailed distribution of recurrent weights.

\subsection{RNN with reciprocal motifs} \label{sec:reciprocal motifs}

The models examined thus far assume that recurrent weights are mutually independent. However, this assumption is not always valid in real-world complex networks. In the brain, for example, certain connectivity motifs are observed more frequently than expected from random connections, which significantly impacts network dynamics~\cite{Shao2025}. Although neuronal correlations in such networks have yet to be studied, it is reasonable to assume that the prevalence of specific connectivity patterns also affects the strength of neuronal correlations. Therefore, we next investigate whether our findings on the relationship between neuronal correlations and memory capacity generalize to RNNs that incorporate such connectivity motifs.

The RNN model studied here is nearly identical to that described in Sec.~\ref{sec2}. The primary difference is that the recurrent weight matrix is now defined as $gJ$, where $g>0$ is a gain parameter. The elements of the matrix $J$ are drawn from a Gaussian distribution with zero mean, $1/N$ variance, and a specific correlation structure given by $[J_{ij} J_{ji}]_J = \eta/N$, following the approach of Marti et al.~\cite{Marti2018}. The parameter $\eta \in [-1,1]$ controls the degree of (anti-)symmetry of connections, i.e., the prevalence of reciprocal motifs. The model reduces to the canonical random RNN when $\eta=0$, while $\eta=1$ and $\eta=-1$ correspond to fully symmetric and fully anti-symmetric connectivity matrices, respectively. It should be noted that we implement this RNN in a discrete-time model, unlike the original continuous-time formulation by Marti et al.~\cite{Marti2018}, because memory capacity is not yet well-defined for continuous-time systems.

In the absence of inputs and noise, the dynamics of the RNN can be analyzed through the eigenvalue spectrum of the recurrent weight matrix $gJ$. According to the Elliptic Law from random matrix theory, the eigenvalues are uniformly distributed over an elliptical disk in the complex plane, with real and imaginary semi-axes of length, $g(1+\eta)$ and $g(1-\eta)$, respectively~\cite{Sommers1988}. Therefore, the trivial fixed point at zero activity, $\bm{x}=\bm{0}$, is stable if and only if this entire eigenvalue spectrum lies within the unit circle. Otherwise, the system can potentially transition to other dynamical regimes, such as a non-zero fixed point, limit cycle, or chaos. While the phase diagram for the continuous-time version of this model was characterized by Marti et al.~\cite{Marti2018}, the corresponding diagram for our discrete-time model is currently unknown. In fact, our own simulations suggest that its unstable regime contains a rich variety of dynamics, including complex oscillations and chaos (data not shown). Therefore, to ensure a clear interpretation of the results, we confine our numerical simulations to the parameter regime where the zero fixed point is stable. Specifically, we fix $g=0.6$ and vary $\eta$ in the range $-2/3 < \eta < 2/3$, which guarantees $g(1 \pm \eta)<1$ is satisfied.  

\subsubsection{Autocorrelation function}

\begin{figure*}[htbp]
    \centering
    \includegraphics[width=0.8\linewidth]{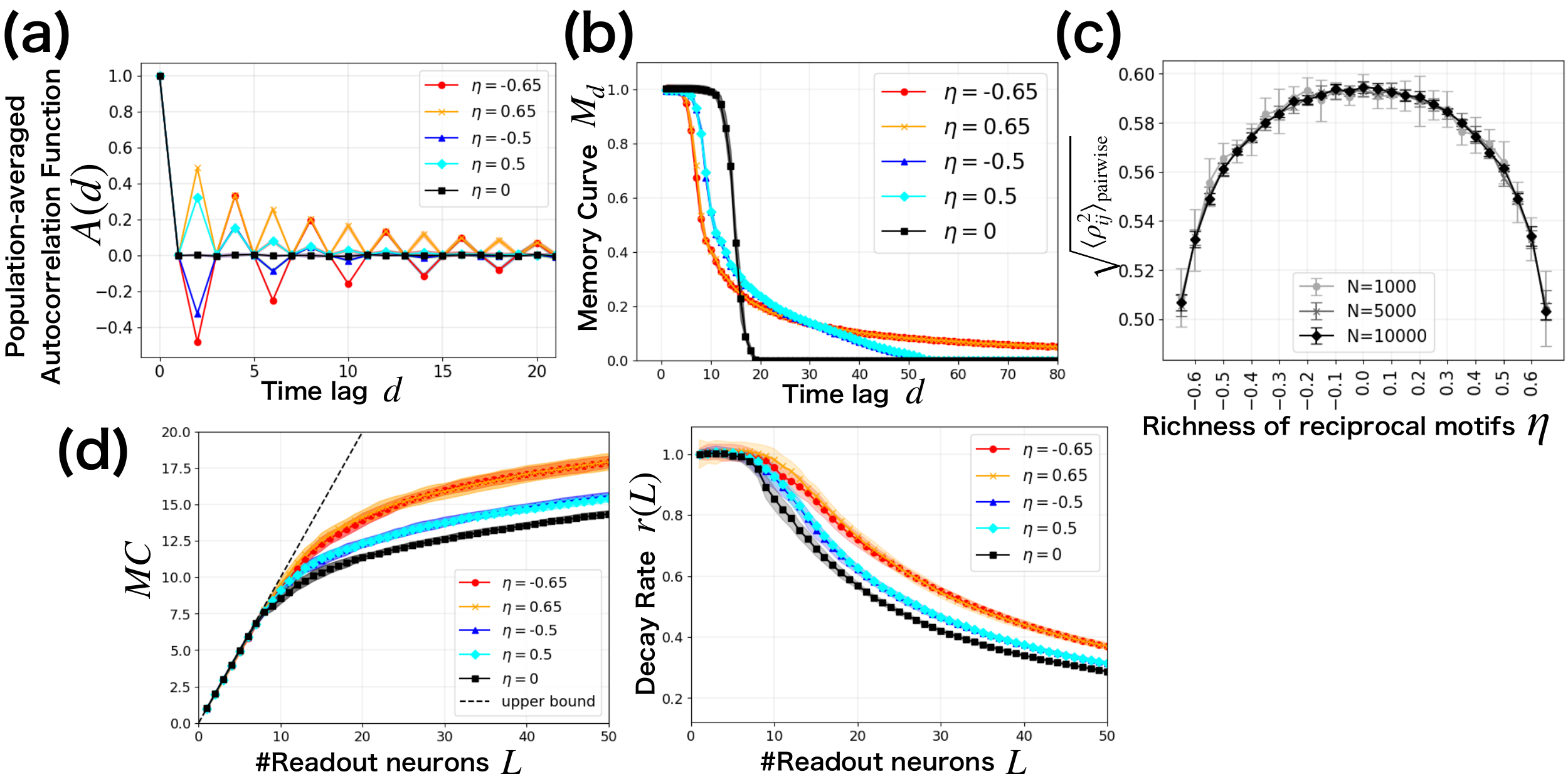}
    \caption{Dynamical statistics and memory capacity of RNNs with reciprocal motifs. (a) Population-averaged autocorrelation function versus time lag $d$. A larger $|\eta|$ results in activity with a longer timescale. (b) Memory function $M_d$ versus time lag $d$. A network with larger $|\eta|$ has better memory for long time lags $d$ but worse memory for short ones. (c) Root means square (RMS) of pairwise Pearson correlation coefficients as a function of $\eta$. (d) Left: Memory capacity as a function of the number of readout neurons, $L$. Right: Decay rate as a function of $L$. Networks with larger $|\eta|$ exhibit weaker neuronal correlations and a correspondingly slower decay of $r(L)$.
    All numerical simulations are performed with input intensity $\sigma_s^2=0.1^2$, no neuronal noise $\sigma_n^2=0$, activation function $\phi=\tanh$, and simulation time length $T=10^4$. The network size is $N=10000$ for all panels except (c), which illustrates the results for $N=1000,5000,10000$. Errorbars and shaded areas represent mean$\pm$std over 10 network and input realizations.
    }
    \label{fig:correlated}
\end{figure*}

First, we examine how the autocorrelation function of our model changes as a function of $\eta$. To do this, we numerically calculate the population-averaged autocorrelation function defined as $A(d) = \frac{1}{N}\sum_i^N \frac{\langle x_i(t)x_i(t-d)\rangle}{\langle x_i(t)^2 \rangle}$. As shown in Fig.~\ref{fig:correlated}(a), an RNN with a larger $|\eta|$ shows a stronger autocorrelation, $| A(d) |$. In contrast, the canonical random model ($\eta=0$) shows a vanishing autocorrelation for all time lags $d\geq 1$. This indicates that a higher density of reciprocal motifs promotes longer timescales in neuronal activity, a finding consistent with continuous-time models~\cite{Marti2018}. A key difference from the continuous-time model, however, is that $A(d)$ is non-zero only for even time lags and vanishes for odd $d$. Intuitively, this is because a perturbation to neuron $i$ at time $t$ can propagate back to itself at time $t+2$ via a two-step path ($i \to j \to i$). The activity at time $t$ is therefore correlated with the activity at $t+2$ due to the underlying correlation between the synaptic weights $J_{ij}$ and $J_{ji}$. It is possible to derive an analytical expression for $A(d)$ using the dynamical cavity method, but we leave this investigation for future work.

\subsubsection{Memory function}

An RNN with long-timescale activity is expected to be better at retaining inputs from the distant past. To test this prediction, we calculate $M_d$ (defined in Eq.(\ref{Md})) for all $d$, which is sometimes referred to as a {\it memory function}. Consistent with our prediction, a larger $| \eta |$ value, which yields a stronger autocorrelation, improves the recall of distant inputs, as shown in Fig.~\ref{fig:correlated}(b). However, these same RNNs perform worse at precisely recalling recent inputs, as indicated by their lower $M_d$ values for short time lags. This happens because the long intrinsic neuronal timescales impose a slow timescale on the reservoir's output, preventing it from tracking rapidly changing Gaussian white noise inputs. We believe this trade-off between recent and distant memory is a general principle in reservoir computing. Indeed, a similar effect was reported in a previous study on a plastic spiking neural network reservoir (see Fig.7(a) in \cite{Cramer2020}). It is important to note that memory capacity, the primary focus of this study, is the sum of $M_d$ over all $d$, and thus, a longer neuronal timescale does not necessarily guarantee a higher memory capacity.

\subsubsection{Neuronal correlations}

Before assessing memory capacity, we examine whether the parameter $\eta$ affects the strength of neuronal correlations. Fig.~\ref{fig:correlated}(c) demonstrates that as $|\eta|$ increases, the strength of neuronal correlations actually decreases. This result may seem counter-intuitive. However, we confirmed that this observation is not a finite-size artifact, as the RMS correlation value remains constant for increasing network sizes $N$. Investigating whether this result is generalizable to other parameter sets and exploring its underlying mechanisms are promising directions for future work, potentially using the dynamical cavity method.

\subsubsection{Scaling behavior of memory capacity}

Given our observation that $|\eta|$ determines the level of neuronal correlations, our hypothesis suggests that it would also influence the decay rate of memory capacity. Our numerical simulations corroborate this prediction. As shown in Fig.~\ref{fig:correlated}(d), larger values of $|\eta|$, which produce weaker neuronal correlations, lead to a slower decrease in the decay rate $r(L)$. Notably, the value of $\eta$ has no effect on the memory capacity of a single neuron, i.e., $MC(L=1)$, while its impact becomes increasingly significant as the number of readout neurons grows. Therefore, the prevalence of reciprocal motifs affects the computational ability of the network as a whole, rather than that of individual neurons, by modulating the structure of neuronal correlations.

\section{Scaling behavior of nonlinear computational capability}

While our previous analysis has focused on memory capacity, non-linear computational ability is equally crucial for RC to address complex tasks. As established in previous works, there exists a trade-off between memory capacity and nonlinear computational ability for RC ~\cite{Dambre2012,Inubushi2017}. Thus, we anticipate that while the growth in memory capacity decelerates as $L$ increases, the system's capacity for nonlinear computation shows a corresponding acceleration.

Information Processing Capacity (IPC) introduced by Dambre et al.~\cite{Dambre2012} offers a task-independent metric for evaluating the performance of RC, enabling a comprehensive assessment of both linear and non-linear computational capabilities. Since IPC theory is somewhat complicated, we provide detailed explanations in Appendix~\ref{appendixC1}. Put simply, the {\it IPC for degree} $D$, denoted by $IPC_D$, represents the reservoir's ability to approximate $D$-th order polynomial functions of past white noise input signals. The IPC has two key properties. First, by definition, memory capacity is exactly equivalent to $IPC_1$, and thus, $IPC_D$ for $D\geq 2$ represents non-linear computational ability. Second, it has been proven that the {\it total IPC}, defined as $IPC_{\rm total} \equiv \sum_{D=1}^\infty IPC_D$, equals to the number of readout units, $L$, provided that the reservoir is a {\it fading memory} dynamical system~\cite{Dambre2012}. A fading memory dynamical system is defined as a system whose state is uniquely determined solely by its input signals~\cite{Dambre2012}.  For example, a reservoir with chaotic dynamics or one subject to noise violates the fading memory property, because its state is determined not only by input signals but also by its initial state or by noise.

\begin{figure*}[htbp]
    \centering
    \includegraphics[width=0.8\linewidth]{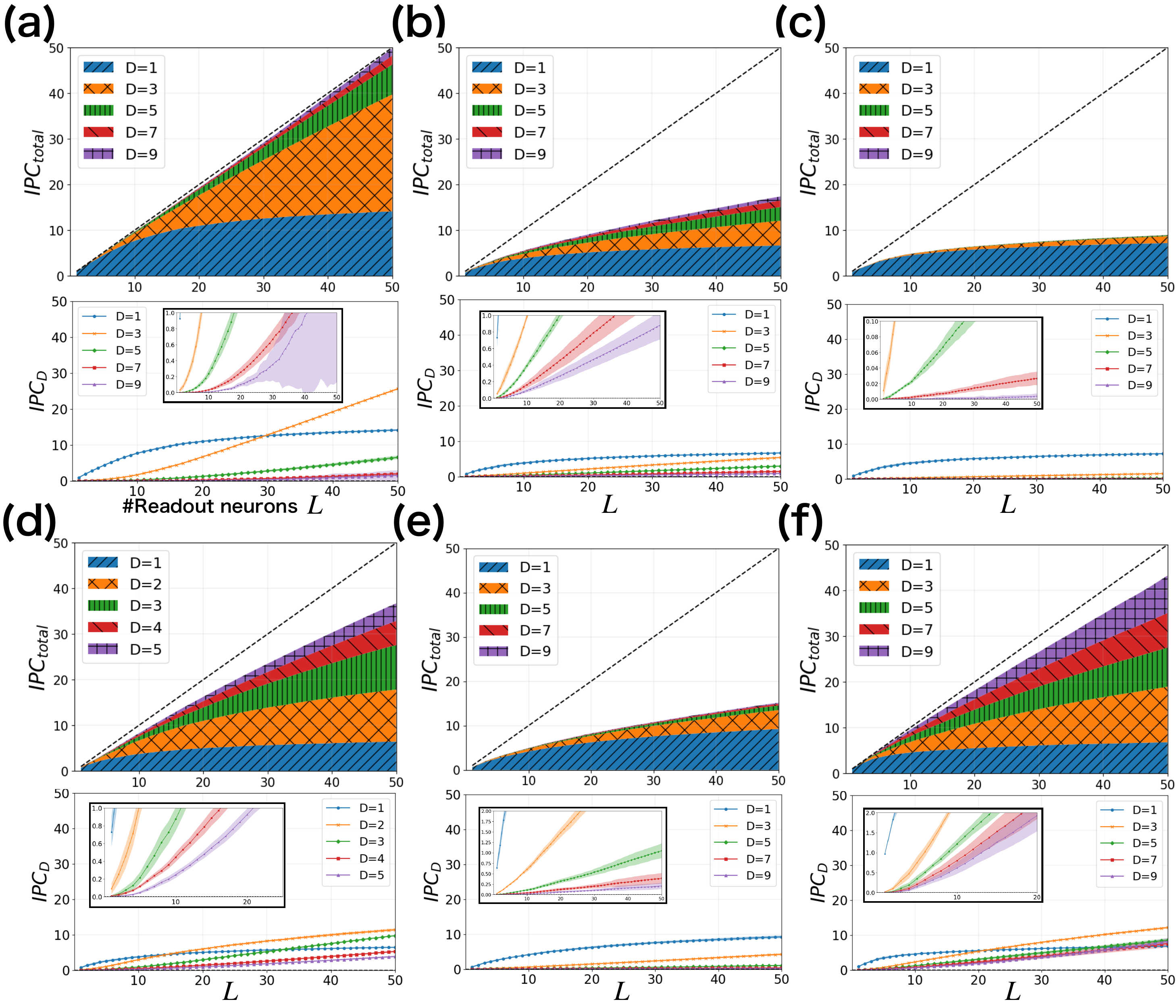}
    \caption{ 
    Scaling behavior of Information Processing Capacity ($IPC_D$) for various dynamical regimes and network architectures. In each main panel (a-f), the top plot shows a stacked representation of the $IPC_D$ values for each degree $D$, while the bottom plot the individual $IPC_D$ values as separate curves. The insets in the bottom plots provide an enlarged view of the smaller IPC values. Overall, the non-linear capacity ($IPC_D$ for $D\geq 2$) arises supralinearly, and the capacities are sequentially activated from lower to higher degrees $D$.
    These capacity values are computed for degrees up to $D=9$, except in panel (d), where it is computed up to $D=5$. (a) Canonical random RNN satisfying the fading memory property ($g=0.9$, $\sigma_s^2=0.3^2$, $\sigma_n^2=0$). (b) Canonical random RNN in a chaotic regime ($g=2.5$, $\sigma_s^2=2.0^2$, $\sigma_n^2=0$). (c) Canonical random RNN with neuronal noise ($g=0.8$, $\sigma_s^2=0.3^2$, $\sigma_n^2=0.1^2$). The activation function for (a-c) is $\phi=\tanh$. (d) Random RNN with a ReLU activation function ($g=1.0$, $\sigma_s^2=1.0$, $\sigma_n^2=0$). In contrast to the other cases, the $IPC_D$ for even $D$ does not vanish due to the asymmetry of the ReLU function. (e) Cauchy RNN with a $\tanh$ activation function ($\gamma=2.0$, $\sigma_s^2=1.0$, $\sigma_n^2=0$). (f) RNN with reciprocal motifs and a $\tanh$ activation function ($g=0.6$, $\eta=0.5$, $\sigma_s^2=1.0$, $\sigma_n^2=0$). 
     In panels (b), (c), and (e), the total measured capacity, $\sum_{D=1}^9 IPC_D$, falls significantly short of the theoretical limit $L$ (dashed line). In contrast, the sum in panel (a) closely matches $L$. This discrepancy is attributed to a violation of the fading memory property. Specifically, the networks in (b) and (e) are chaotic, as confirmed by their positive maximum conditional Lyapunov exponents~\cite{Pikovsky2016,Takasu2024}  ($\lambda_{\rm max}=0.089\pm 0.0041$ and $\lambda_{\rm max}=0.23\pm0.011$, respectively); meanwhile, the network in (c) is subject to noise. Note that the networks in panels (d) and (f) satisfy the fading memory property, but the sums are slightly less than $L$ due to the calculation being truncated at a finite degree ($D\leq5$ and $D\leq9$, respectively). Shaded areas represent mean$\pm$std for 10 different network and input signal realizations. All simulations are performed with a network size of $N=1000$ and a simulation time of $T=10^5$. The transient washout time is $T_{\rm washout}=10^3$.
    }
    \label{fig:ipc}
\end{figure*}

Fig.\ref{fig:ipc}(a)-(c) illustrate how the values of $IPC_D$ of the canonical random RNN, defined in Sec.~\ref{sec2}, change with increasing $L$ across various hyperparameter combinations ($g$, $\sigma_s^2$, and $\sigma_n^2$), as obtained through numerical simulations. Note that, due to the model's symmetry, $IPC_D$ values for all even degrees $D$ are identically zero. As shown, a decline in the growth of the $IPC_1$ (equivalent to memory capacity) is accompanied by an increase in the non-linear computational ability (represented by $IPC_D$ for $D\geq 3$), reflecting memory-nonlinearity trade-off. 

Our novel finding is that, as $L$ increases, the values of higher-degree $IPC_D$ are activated {\it sequentially} from lower to higher orders, accompanied by their {\it supralinear} rises. This distinctive scaling behavior of nonlinear computational ability might be considered a trivial result in RNNs that satisfy the fading memory property, where the total capacity is constrained by $\sum_{D\geq 1} IPC_D=L$. In such cases, a sublinear growth in memory capacity ($=IPC_1$) necessarily implies that the remaining capacity grows supralinearly. However, the significance of our finding lies in its generality. This sequential, supralinear arising of nonlinear computational powers is observed not only in the RNNs that satisfy the fading memory property (Fig.~\ref{fig:ipc}(a)), but also in chaotic (Fig.~\ref{fig:ipc}(b)) and noisy (Fig.~\ref{fig:ipc}(c)) RNNs, where this property breaks down and thus the equality $MC+\sum_{D\geq 2}IPC_D =L$ does not hold. Furthermore, the phenomenon holds across various network architectures, including those with a ReLU activation function (Sec.~\ref{sec:relu}), heavy-tailed weight distributions (Sec.~\ref{sec:cauchy}), and reciprocal connectivity motifs (Sec.~\ref{sec:reciprocal motifs}), as shown in Fig.~\ref{fig:ipc}(d)-(f). 

As analytical derivation of higher-degree $IPC_D$ is beyond our capability, the detailed mechanism underlying these observations remains unclear. However, it is plausible that the neuronal correlation plays a pivotal role, analogous to its influence on memory capacity. This hypothesis is supported by speculation that each $IPC_D$ would exhibit linear scaling with $L$ if the effects of neuronal correlations were neglected, which contrasts with the observed supralinear and sequential emergence of $IPC_D$ (see Appendix~\ref{appendixC2}).

\section{Discussion}

In the present study, we investigate how the computational capacity of large random RNN reservoirs scales as increasing the number of readout neurons, $L$. We develop a theoretical framework for deriving memory capacity that can incorporate the effect of neuronal correlations, which has been ignored for analytical simplicity. Our theory reveals that memory capacity scales {\it sublinearly} with $L$, and the extent of this sublinearity is significantly influenced by the strength of neuronal correlations, even though those correlations are quite weak. Although our theory is limited to cases where neuronal correlations scale as $O(1/\sqrt{N})$, a constraint  imposed by specific assumptions about our model's hyperparameters (Eq.(\ref{scaling})), our numerical simulations confirm the robustness of this finding across a wider range of hyperparameters and in various network architectures. In addition, we numerically study the scaling behavior of nonlinear computational ability, showing that higher-order computational powers emerge {\it successively}, accompanied by their {\it supralinear} rises, enabling the reservoir to handle increasingly complex tasks. 

The specific memory and nonlinearity requirements of a given task can be assessed using the framework proposed by Hülser et al.~\cite{Hulser2023}. In this context, our results provide a practical principle for designing cost-effective RC systems: while increasing L raises computational and structural costs and yields diminishing returns for memory, it becomes essential when tasks demand higher-order nonlinear computations.

Beyond its practical usefulness, our work offers a new lens for understanding RC performance. The computational ability of RC has been predominantly discussed from the perspective of proximity to the edge of chaos~\cite{Bertschinger2004, Legenstein2007, Schuecker2018, Haruna2019}, typically for a fixed number of readout neurons~\cite{Dambre2012, Inubushi2017}. To our knowledge, this is the first study to analyze the computational ability from the novel perspective of its scaling behavior with an increasing number of readout neurons and the corresponding role of neuronal correlations.  Interestingly, a recent report has suggested that the edge of chaos is not always optimal~\cite{Carroll2020}, but the reason for this remains unclear. The new perspective we propose may hold a crucial key to unraveling this puzzle.

It is also instructive to contrast our findings with other models of memory, such as Hopfield networks~\cite{Hopfield1982, Amit1985}.  In the field of Hopfield networks, it is well-established that making network activity sparse dramatically improves the capacity of associative memory~\cite{Amari1989, Okada1996}. Since sparse activity tends to weaken correlations between neurons~\cite{Vinje2000}, this implies that the strength of neuronal correlations is strongly associated with the memory ability of Hopfield networks. At first glance, our work might seem to merely replicate these prior findings. However, there is a fundamental difference between them. The mechanism by which Hopfield networks store information utilizes the fixed points of attractor dynamics, which can be described as static and long-term memory. RC, in contrast, takes the approach of temporarily retaining time-series information through reverberating neural activity within a reservoir network, which is a form of dynamic and short-term memory (or working memory). Given that their memory mechanisms are entirely different, there is no guarantee that a principle governing static memory would apply equally to dynamic memory. 

Our findings have implications that extend beyond the framework of RC. First, in neuroscience, the mammalian cortex in the awake state typically operates in an asynchronous regime characterized by near-zero neuronal correlations~\cite{Vreeswijk1996,Renart2010, Dahmen2019}, while recent studies have shown the level of correlation is structured~\cite{Rosenbaum2018, Safavi2018} and influenced by factors such as stimulus drive~\cite{Churchland2010,Tan2014} and attention~\cite{Cohen2009, Mitchell2009}. The functional impacts of the neuronal correlations have been extensively investigated through the lens of neural manifolds~\cite{Fusi2016, Stringer2019, Farrell2022}. Within this framework, a neuronal network with a high level of correlated neuronal activity occupies a restricted region of $N$-dimensional phase space, the linear dimension of which is quantified by the effective dimension~\cite{Rajan2010}. In the context of population coding, a well-established trade-off exists between representational richness and robustness: high-dimensional (i.e., low-correlation) activity enables complex and diverse information encoding, while low-dimensional (i.e., high-correlation) activity supports redundant encoding and noise robustness~\cite{Stringer2019, Farrell2022}. Existing theoretical frameworks, however, have primarily focused on the implications of neuronal correlations for static information processing, i.e., mapping static inputs to outputs. Our findings suggest a novel role for neuronal correlations in dynamic information processing, offering a new hypothesis regarding their functional significance in the brain.

Second, in the field of deep learning theory, it has been established that neural networks trained via gradient-based methods, such as BPTT, can learn tasks both in the {\it lazy regime}, where their weights undergo only infinitesimal changes from their initial random values~\cite{Jacot2018,Lee2019}, and in the {\it rich regime}, where weights change substantially to extract task-relevant features~\cite{Yang2021, Bordelon2022, Fischer2024}. Our insights may provide valuable understanding of the computational capacity of RNNs trained in the lazy regime, as their recurrent weights remain nearly identical to those of untrained, reservoir-style networks.

In conclusion, our study not only advances the understanding of RC but also sheds light on the broader implications of neuronal correlations and their effects on both biological and artificial systems. These insights offer new directions for optimizing RNN architectures and exploring the fundamental nature of computation in complex networks.

{\it Data availability}: The source code used to generate all figures, except for Fig.\ref{fig:schematic of DCM}, is available at \cite{source_code}.

\begin{acknowledgments}
We thank Joni Dambre for providing the source code for computing information processing capacity, and Tomoyuki Kubota for helpful discussions regarding its implementation. S.T. was supported by JSPS KAKENHI Grant No. JP22KJ1959. T.A. was supported by MEXT KAKENHI Grant Numbers 23H04467 and by JSPS KAKENHI Grant Numbers 24H00723, 20K20520.
\end{acknowledgments}

%


\begin{widetext}

\appendix

\section{THEORETICAL ANALYSIS ON STATISTICS OF NEURONAL ACTIVITY}
\label{appendixA}
In this section, we derive statistics of neuronal activity based on dynamical cavity approach~\cite{Clark2022, Zou2024}. Several results and methodologies introduced here are frequently utilized in deriving memory capacity in Appendix~\ref{appendixB}. 

In deriving single neuronal statistics such as the variance of $x_i(t)$ (see Appendix~\ref{appendixA1}) in the limit of $N\to\infty$, the same result can be obtained via dynamical mean-field theory~\cite{Sompolinsky1988, Helias} . However, since the dynamical mean-field theory transforms the $N$-body system into the effective single-body system, it is inherently unable to evaluate many-body neuronal statistics, such as neuronal correlations (see Appendix~\ref{appendixA2}) and memory capacity(see Appendix~\ref{appendixB1}). In contrast, the dynamical cavity method is, in principle, applicable to $n$-body neuronal statistics for any $n$. Furthermore, while the aforementioned limitation of dynamical mean-field theory has recently been addressed through subleading corrections to the saddle-point solution~\cite{Clark2024, Dick2024}, the dynamical cavity method remains advantageous for its conceptual clarity. 

In the following, we use the shorthand notation for brevity $x_{i,t}\equiv x_i(t)$, $s_t \equiv s(t)$, $\phi_{i,t}\equiv \phi(x_{i,t})$ and $\langle x_i x_j \rangle \equiv \langle x_i(t) x_j(t) \rangle $.

Before discussing the details, we first outline the necessity of employing the dynamical cavity method. Consider calculating the quenched variance of neuronal activity, $[x_{i,t}^2]$. For simplicity, we ignore input and noise signals. According to time evolution equation of neurons (Eq.(\ref{model})), $x_{i,t}$ is the sum of a large number ($N\gg 1$) of mutually independent quantities, $J_{ij} \phi(x_{j,t-1})$. This implies that $x_{i,t}$ follows a Gaussian distribution due to the central limit theorem, and thus, it is sufficient to determine the complete statistics of $x_{i,t}$ by calculating its mean and variance. Given the symmetry of the model, we have $[ x_{i,t}^2]=0$. On the other hand, we can intuitively calculate $[ x_{i,t}^2]$ as:
\begin{eqnarray} \label{intuitive calc}
    [ x_{i,t}^2]= \sum_{j,k}^N \left[J_{ij}J_{ik}  \phi_{j,t-1} \phi_{k,t-1}  \right] 
    \overset{?}{=} \sum_{j,k}^N \left[J_{ij}J_{ik} \right] \left[  \phi_{j,t-1} \phi_{k,t-1}  \right] 
    = g^2 \left[ \phi_{j,t-1}^2 \right]
    = g^2 \langle \phi(x)^2 \rangle_{x\sim \mathcal{N}(0, [x_{i,t-1}^2])}, 
\end{eqnarray}
yielding time series of $[x_{i,t}^2]$. Here we assumed in the second equation that the recurrent weights, $J_{ij}$ and $J_{ik}$, are independent of the neuronal activities, $\phi_j$ and $\phi_k$. Fortunately, this assumption is validated in the limit of $N\to\infty$, as shown with the dynamical mean-field theory. However, this assumption is likely to break down when finite size effects are taken into account. To circumvent this challenge, we need to employ the dynamical cavity method.

\begin{figure}[htbp]
    \centering
    \includegraphics[width=0.4\linewidth]{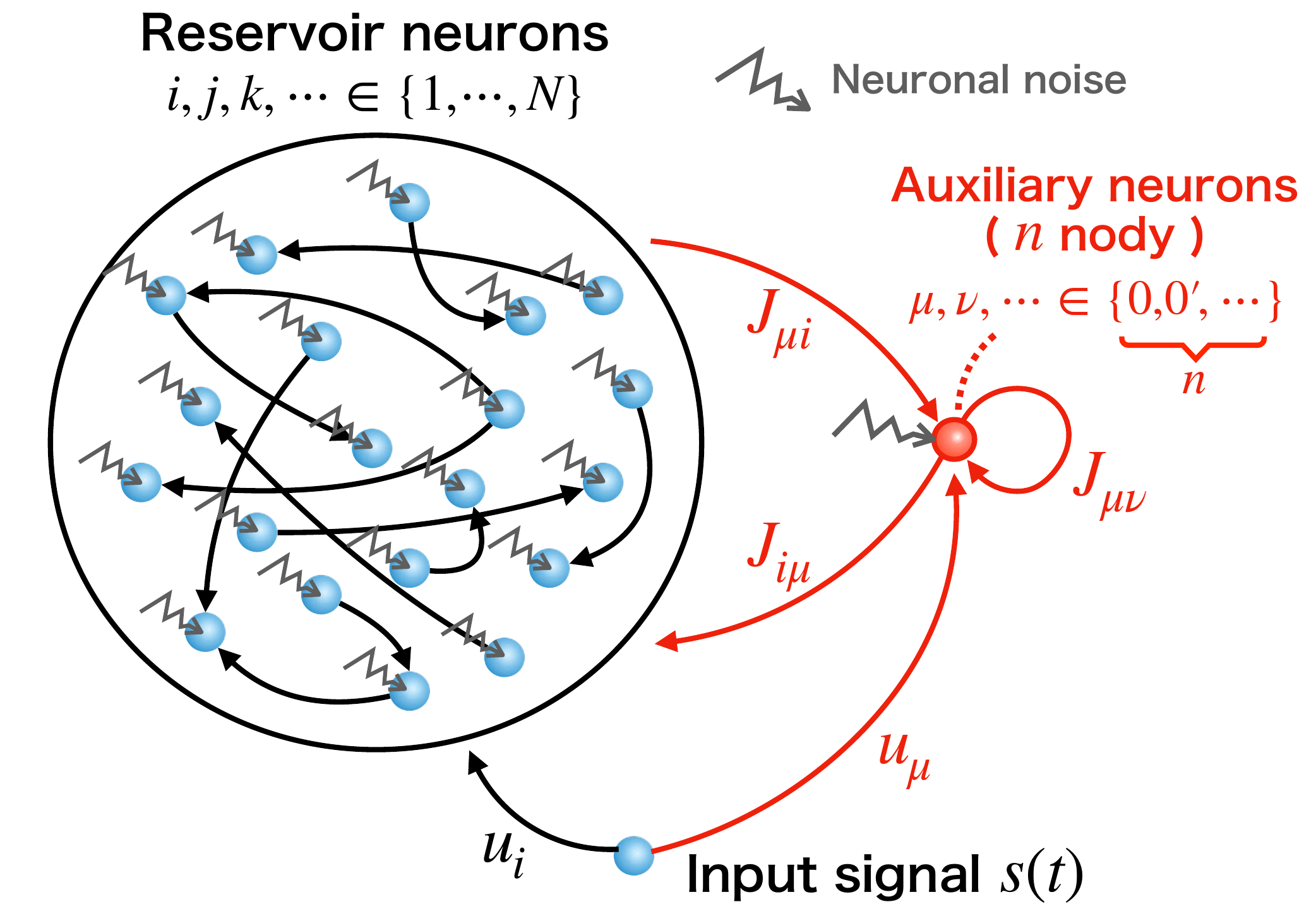}
    \caption{ 
    Overview of dynamical cavity method. For simplicity, only one auxiliary neuron is illustrated, but in practice, $n$ auxiliary neurons are introduced to calculate $n$-body statistics. Latin indices, $i,j,\cdots$, are used to label reservoir neurons, while Greek indices, $\mu, \nu, \cdots$, denote auxiliary neurons. The weights connecting auxiliary neurons to reservoir neurons, as well as the self-connections of auxiliary neurons, are sampled from the same distribution as the recurrent weights of the reservoir. 
    }
    \label{fig:schematic of DCM}
\end{figure}

\subsection{Single neuronal statistics} \label{appendixA1}

We aim to derive the quenched average of the time-variance of neuronal activity, $[\langle x_i^2 \rangle]$. Note that due to the self-averaging, it is equivalent to the population average of time-variance to leading order, i.e. $[\langle x_i^2 \rangle] = \frac{1}{N} \sum_i^N \langle x_i^2 \rangle$. The symmetry of the model setting ensures that the quenched average of the time-mean, $[\langle x_i \rangle]$, vanishes.

The overview of dynamical cavity method is described in Fig.\ref{fig:schematic of DCM}. One auxiliary neuron, indexed by $0$, is added to the original reservoir RNN whose neurons are indexed by $i=1,\cdots, N$. The original neuron is perturbated by the addition of the auxiliary neuron. By denoting the perturbation applied to the $i$th neuron at time $t$ by $\delta_{i,t}$, we describe the dynamics of the auxiliary neuron as
\begin{eqnarray} \label{1body auxiliary dynamics}
x_{0,t+1} &&= \sum_{i=1}^N J_{0i}\phi(x_{i,t} + \delta_{i,t}) + J_{00}\phi_{0,t} + u_0 s_{t+1} + \xi_{0,t+1} \nonumber \\
&&= \sum_{i=1}^N J_{0i}\phi_{i,t} + \sum_{i=1}^N J_{0i}\phi'_{i,t} \delta_{i,t}+ J_{00}\phi_{0,t} + u_0 s_{t+1} + \xi_{0,t+1}.
\end{eqnarray}
Here, it should be noted that $x_{i,t}$ represents the original activity of a reservoir neuron before the introduction of the auxiliary neuron. Therefore, the value of $J_{0i}$ is completely independent of reservoir activities, $x_{i,t}$ and $\phi_{i,t}$, in contrast to the intuitive argument in Eq.(\ref{intuitive calc}). 

We define $\chi_{ij,t\ t-\tau}\equiv \delta x_{i,t}/\delta I_{j,t-\tau}$ as the linear susceptibility of the $i$th neuron at time t, induced by an infinitesimal external inputs applied to the $j$th neuron at time $t-\tau \ (\tau\geq 1)$, $I_{j,t-\tau}$. Then, as the perturbation to $i$th neuron at time $t$, $\delta_{i,t}$, is the sum of all direct perturbations to $j\in \{1,\cdots,N\}$ at time $t-\tau\ (\tau \geq 1)$ weighted by the corresponding susceptibility $\chi_{ij,t\ t-\tau}$, we obtain
\begin{eqnarray} \label{perturbation}
\delta_{i,t} = \sum_{\tau \geq 1} \sum_{j=1}^N \chi_{ij,t\ t-\tau} J_{j0}\phi_{0,t-\tau-1}. 
\end{eqnarray}
We substitute Eq.(\ref{perturbation}) into Eq.(\ref{1body auxiliary dynamics}), yielding
\begin{eqnarray}\label{AtoD dynamics}
x_{0,t+1} &&=  \underbrace{\sum_{i=1}^N J_{0i}\phi_{i,t}}_{\equiv \eta_t} + \sum_{\tau \geq 1}\underbrace{\sum_{i,j}^N \chi_{ij,t\ t-\tau}J_{0i}J_{j0}\phi'_{i,t}}_{\equiv\kappa_\tau} \phi_{0,t-\tau-1} + J_{00}\phi_{0,t}+ u_0 s_{t+1} + \xi_{0,t+1} \nonumber\\
&&\equiv \eta_t + \sum_{\tau \geq 1} \kappa_\tau \phi_{0,t-\tau-1}+ J_{00}\phi_{0,t}+ u_0 s_{t+1} + \xi_{0,t+1}. \nonumber\\
\end{eqnarray}
The physical interpretation of this equation is as follows. The first term, $\eta_t$, is the feedforward input from the reservoir. The second term, $\kappa_\tau \phi_{0,t-\tau-1}$, arises from the perturbation induced in the reservoir by the auxiliary neuron $0$, which is read out by neuron $0$. The third term originates from self-connection of the auxiliary neuron $0$.  

The order of the first term is evaluated as
\begin{eqnarray} \label{order of term 1}
\left[\eta_t^2\right] = \sum_{i,j}^N [J_{0i}J_{0j} \phi_{i,t}\phi_{j,t}] = \sum_{i,j}^N [J_{0i}J_{0j}][\phi_{i,t}\phi_{j,t}] = g^2[\phi^2_{i,t}] \sim O(1) 
\end{eqnarray}
Here we used in the second equation the fact that $J_{0i}$ and $J_{0j}$ are independent of $\phi_{i,t}$ and $\phi_{j,t}$ as mentioned above. The order of the third term is evaluated as
\begin{eqnarray} \label{order of term 3}
\left[\left(J_{00}\phi_{0,t} \right)^2\right] \leq [J_{00}^2] \sim O(1/N).
\end{eqnarray}

For evaluation of the second term in Eq.(\ref{AtoD dynamics}), we assess the order of $\chi_{ij, t\ t-\tau}$. We begin with $\tau=1$. Since $\chi_{ij,t\hspace{1pt}t-1} = J_{ij}\phi'_{j,t-1}$, we find  $[\chi_{ij,t\hspace{1pt}t-1}^2 ] \leq [J_{ij}^2] \sim O(1/N)$. Subsequently, for $\tau=2$,  $\chi_{ij,t\hspace{1pt}t-2} = \sum_k^N J_{ik}\phi'_{k,t-1}\chi_{kj,t-1\hspace{1pt}t-2}$, leading to $[\chi_{ij,t\hspace{1pt}t-2}^2 ] \leq \sum_k^N [J_{ik}^2][(\chi_{kj,t-1\hspace{1pt}t-2})^2] \sim O(1/N)$, where we used the result for $\tau=1$, $[(\chi_{kj,t-1\hspace{1pt}t-2})^2]\sim O(1/N)$. By inductively repeating this analysis, we conclude that $[\chi_{ij,t\ t-\tau}^2] \sim O(1/N)$ for any $\tau \geq 1$. Therefore, the order of the second term is evaluated as 
\begin{eqnarray}\label{order of term 2}
[\kappa_\tau^2] &&= \sum_{i,j,k,l}^N [J_{0i}J_{j0}J_{0k}J_{l0}][\phi'_{i,t}\phi'_{k,t}\chi_{ij,t\ t-\tau}\chi_{kl,t\ t-\tau}] 
\leq \frac{g^4}{N^2}\sum_{i,j}^N [\chi_{ij,t\ t-\tau}^2] \sim O(1/N).
\end{eqnarray}
Eventually, we obtain $[\kappa_\tau^2] \sim O(1/N)$ for $\tau \geq 1$. 

Employing Cauchy–Schwarz inequality, it is shown that all cross terms between $\eta_t$, $\kappa_\tau$, and $J_{00}\phi_{0,t}$ are $O(1/\sqrt{N})$.

Leveraging all these order estimates, from Eq.(\ref{AtoD dynamics}), we obtain
\begin{eqnarray}
[x_{0,t+1}^2] &&= \left[\big(\eta_t + \sum_{\tau \geq 1} \kappa_\tau \phi_{0,t-\tau-1}+ J_{00}\phi_{0,t}+ u_0 s_{t+1} + \xi_{0,t+1} \big)^2 \right] \nonumber \\
&&= [\eta_t^2] + [u_0^2]s_{t+1}^2 + [\xi_{0,t+1}^2] + O(1/\sqrt{N}) \nonumber \\
&&= g^2 [\phi_{i,t}^2] + s_{t+1}^2 + \sigma_n^2 +  O(1/\sqrt{N}).
\end{eqnarray}
Thus, to leading order, we can ignore the non-trivial terms arising from the effects of the perturbations induced by the auxiliary neuron. The statistical behavior of the auxiliary neuron is equivalent to that of the reservoir neurons, which enables the replacement of $[x_{0,t}^2]$ with $[x_{i,t}^2]$, resulting in 
\begin{eqnarray}\label{x^2 large input}
[x_{i,t+1}^2] = g^2[\phi_{i,t}^2] + s_{t+1}^2 + \sigma_n^2 + O(1/\sqrt{N}). 
\end{eqnarray}
For the scaling assumption given by Eq.(\ref{scaling}), we can assume the inputs to be subleading, obtaining 
\begin{eqnarray}
[x_{i,t+1}^2] = g^2[\phi_{i,t}^2] + \sigma_n^2 + O(1/\sqrt{N}).
\end{eqnarray}
Taking the time average and the network size limit ($N\to\infty$) yields a {\it{dynamical mean-field equation}},
\begin{eqnarray} \label{DMFeq appendix}
[\langle x_{i}^2 \rangle] = g^2[\langle \phi_{i}^2 \rangle ] + \sigma_n^2,
\end{eqnarray}
which is identical to Eq.(\ref{DMFeq}) in the main text.

For later use, we evaluate the order of the quenched variance of $\langle x_i^2 \rangle$, denoted by ${\rm{Var}}[\langle x_i^2 \rangle]$, which is equivalent to the population variance because of the self-averaging property. Since the cavity method allows us to ignore the effects of perturbation induced by the auxiliary neuron, we easily calculate $[\langle x_0^2 \rangle^2]$ as 
\begin{eqnarray} \label{[<x^2>^2]}
[\langle x_0^2 \rangle^2] &&= \left[ \left\langle \left( \sum_i^N J_{0i}\phi_{i,t-1} + u_0 s_t + \xi_{0,t} \right)^2 \right\rangle^2\right] \nonumber \\
&&= \frac{2g^4}{N^2}\sum_{i,j}^N [\langle \phi_i \phi_j \rangle^2] + \frac{g^4}{N^2}\sum_{i,j}^N [\langle \phi_i^2 \rangle \langle \phi_j^2 \rangle] + 2g^2 (\sigma_s^2 + \sigma_n^2)[\langle \phi_i^2 \rangle] + 3\sigma_s^4 + \sigma_n^4 \nonumber\\
&&= 2g^4 [\langle \phi_i \phi_j \rangle^2]_{i\neq j} + \frac{3g^4}{N}[\langle \phi_i^2 \rangle^2] +g^4[\langle \phi_i^2 \rangle \langle \phi_j^2 \rangle]_{i\neq j} + 2g^2 (\sigma_s^2 + \sigma_n^2)[\langle \phi_i^2 \rangle] + 3\sigma_s^4 + \sigma_n^4.
\end{eqnarray}
Employing Eq.(\ref{x^2 large input}) and Eq.(\ref{[<x^2>^2]}), we obtain
\begin{eqnarray}
{\rm{Var}}[\langle x_0^2 \rangle] &&= [\langle x_0^2 \rangle^2] - [\langle x_0^2 \rangle]^2 \nonumber\\
&&=g^4 \left( [\langle \phi_i^2 \rangle \langle \phi_j^2\rangle]_{i\neq j}-[\langle \phi_i^2 \rangle ]^2\right) +2g^4 [\langle \phi_i \phi_j \rangle^2]_{i\neq j} + \frac{3g^4}{N}[\langle \phi_i^2 \rangle^2] + 2\sigma_s^4.
\end{eqnarray}
Specifically, when $\sigma_s^2 = \tilde{\sigma}_s^2/\sqrt{N}$, we have
\begin{eqnarray} \label{var x^2}
{\rm{Var}}[\langle x_0^2 \rangle] &&=g^4 \left( [\langle \phi_i^2 \rangle \langle \phi_j^2\rangle]_{i\neq j}-[\langle \phi_i^2 \rangle ]^2\right) +2g^4 [\langle \phi_i \phi_j \rangle^2]_{i\neq j} + \frac{3g^4}{N}[\langle \phi_i^2 \rangle^2] + \frac{2 \tilde{\sigma}_s^4}{N}.
\end{eqnarray}
As shown in Appendix~\ref{appendixA2}, we see $[\langle \phi_i \phi_j \rangle^2]_{i\neq j}\sim O(1/N)$. The self-averaging property allows us to evaluate the first term as
\begin{eqnarray}
[\langle \phi_i^2 \rangle \langle \phi_j^2\rangle]_{i\neq j}-[\langle \phi_i^2 \rangle ]^2
\sim \frac{1}{N^2-N}\sum_{i,j\ (i\neq j)}^N \langle \phi_i^2 \rangle \langle \phi_j^2\rangle - \left(\frac{1}{N}\sum_i^N \langle \phi_i^2 \rangle \right)^2 \sim O\left(\frac{1}{N}\right).
\end{eqnarray}
Consequently, we obtain ${\rm{Var}}[\langle x_i^2 \rangle]\sim O(1/N)$.

A direct consequence of the scaling ${\rm{Var}}[\langle x_i^2 \rangle] \sim O(1/N)$ is the following equation to leading order: 
\begin{eqnarray} \label{mean of inverse}
\left[ \frac{1}{\langle x_i^2 \rangle}\right] =  \frac{1}{[\langle x_i^2 \rangle]},
\end{eqnarray}
provided that $[\langle x_i^2 \rangle] \sim O(1)$.

\subsection{Statistics for neuronal correlations} \label{appendixA2}
We aim to derive the quenched average of the squared neuronal correlation, $[\langle x_i x_j \rangle^2]$, which is equivalent to the population average to leading order, $[\langle x_i x_j \rangle^2]=\frac{1}{N^2-N} \sum_{i,j(i\neq j)}^N\langle x_i x_j \rangle^2$. Similarly to the single neuronal statistics, the model's symmetry ensures that $[\langle x_i x_j \rangle] = 0$. 

As in Appendix~\ref{appendixA1}, we introduce two auxiliary neurons indexed by $0$ and $0'$ to the original reservoir RNN. Following the notation in Clark's work~\cite{Clark2022},  Latin indices are used to label reservoir neurons, while Greek indices denote auxiliary neurons. The dynamics of the auxiliary neurons $\mu\in \{0,0'\}$ is described by
\begin{eqnarray} \label{2body dynamics}
x_{\mu,t+1} &&= \sum_i^N J_{\mu i}\phi(x_{i,t}+\delta_{i,t}) + \sum_\nu J_{\mu \nu} \phi_{\nu, t} + u_\mu s_{t+1} + \xi_{\mu,t+1} \nonumber \\
&&= \sum_i^N J_{\mu i}\phi_{i,t} + \sum_i^N J_{\mu i}\phi'_{i,t}\delta_{i,t} +\sum_\nu J_{\mu \nu} \phi_{\nu, t} + u_\mu s_{t+1} + \xi_{\mu,t+1}.
\end{eqnarray}
The perturbation $\delta_{i,t}$ is given by
\begin{eqnarray} \label{perturbation for 2body}
\delta_{i,t} = \sum_{\tau \geq 1}\sum_\nu \sum_{j}^N \chi_{ij,t\ t-\tau}J_{j\nu}\phi_{\nu,t-\tau-1}.
\end{eqnarray}
Substituting Eq.(\ref{perturbation for 2body}) into Eq.(\ref{2body dynamics}), we obtain
\begin{eqnarray}\label{AtoD for two cavity}
x_{\mu,t+1} &&= \underbrace{\sum_i^N J_{\mu i}\phi_{i,t}}_{\equiv \eta_{\mu,t}} + \sum_{\tau \geq 1} \sum_{\nu}\underbrace{\sum_{i,j}^N J_{\mu i}J_{j\nu}\phi'_{i,t} \chi_{ij,t\ t-\tau}}_{\equiv \kappa_{\mu\nu,t\ t-\tau}} \phi_{\nu,t-\tau-1} +\sum_\nu J_{\mu \nu} \phi_{\nu, t} + u_\mu s_{t+1} + \xi_{\mu,t+1} \nonumber \\
&&\equiv \eta_{\mu, t} + \sum_\nu \sum_{\tau \geq 1} \kappa_{\mu\nu,t\ t-\tau}\phi_{\nu,t-\tau-1} + \sum_\nu J_{\mu \nu} \phi_{\nu, t} + u_\mu s_{t+1} + \xi_{\mu,t+1}. \nonumber\\
\end{eqnarray}
The physical interpretation of this equation is analogous to the case of single neuronal statistics (Eq.\ref{AtoD dynamics}): $\eta_{\mu,t}$ denotes the direct feedforward input from the reservoir to the auxiliary neuron $\mu$, and $\kappa_{\mu \nu, t\ t-\tau}\phi_{\nu,t-\tau-1}$ arises from the perturbation induced in the reservoir by the auxiliary neuron $\nu$ at time $t-\tau-1$, which is read out by the auxiliary neuron $\mu$ at time $t$. 

Expanding $[\langle x_0 x_{0'} \rangle^2]$ using Eq.(\ref{AtoD for two cavity}), we need to evaluate all cross terms. We can accomplish this task through direct calculation, analogous to the approach used in Appendix~\ref{appendixA1}. First, 
\begin{eqnarray}
[\langle \eta_{0,t}\eta_{0',t} \rangle^2] &&= \sum_{i,j,k,l}^N \left[ J_{0i}J_{0'j}J_{0k}J_{0'l}\langle \phi_i\phi_j\rangle\langle \phi_k\phi_l\rangle \right] = \sum_{i,j,k,l}^N [J_{0i}J_{0'j}J_{0k}J_{0'l}][\langle \phi_i\phi_j\rangle\langle \phi_k\phi_l\rangle] = \frac{g^4}{N}[\langle \phi_i^2 \rangle^2] + g^4 [\langle \phi_i \phi_j \rangle^2]_{(i\neq j)}, \nonumber\\
\end{eqnarray}
where we used in the second equation the fact that the weights connecting to auxiliary neurons are independent from the reservoir neurons' activities. By making an ansatz, $[\langle \phi_i \phi_j \rangle^2]_{(i\neq j)} \sim O(1/N)$, we obtain $[\langle \eta_{0,t}\eta_{0',t} \rangle^2] \sim O(1/N)$. 

Next, calculating the cross term between the second and third terms yields
\begin{eqnarray}
\left[ \left\langle \sum_{\mu}\sum_{\tau \geq 1} \kappa_{0\mu,t\ t-\tau} \phi_{\mu,t-\tau-1} \sum_\nu J_{0'\nu} \phi_{\nu,t} \right\rangle^2\right] &=& \frac{g^2}{N} \sum_{\mu, \nu, \rho}\sum_{\tau, \tau'} \left[ \langle \kappa_{0\mu, t\ t-\tau} \phi_{\mu,t-\tau-1} \phi_{\nu, t} \rangle \langle \kappa_{0\rho, t\ t-\tau} \phi_{\rho,t-\tau'-1} \phi_{\nu, t} \rangle  \right] \\
&=& \frac{g^2}{N} \sum_{\mu, \nu, \rho}\sum_{\tau, \tau'} \sum_{i,j,k,l}^N [J_{0i}J_{j\mu}J_{0k}J_{l\nu}] [\langle \phi'_{i,t}\chi_{ij, t\ t-\tau} \phi_{\mu,t-\tau-1}\phi_{\nu,t} \rangle \nonumber\\
&& \hspace{120pt}\times \langle \phi'_{k,t}\chi_{kl, t\ t-\tau'} \phi_{\rho,t-\tau'-1}\phi_{\nu,t} \rangle ] \nonumber \\
&=& \frac{g^6}{N^3}\sum_{\mu}\sum_{\tau, \tau'} \sum_{i,j}^N
 [\langle \phi'_{i,t} \rangle^2 \langle \chi_{ij, t\ t-\tau} \chi_{ij, t\ t-\tau'}\rangle \langle \phi_{\mu,t-\tau-1}\phi_{\mu,t} \rangle \langle \phi_{\mu,t-\tau'-1}\phi_{\mu,t} \rangle ] \nonumber \\
 &\sim& O(1/N^2),
\end{eqnarray}
where we used $\chi_{ij, t\ t-\tau}^2 \sim O(1/\sqrt{N})$ for any $\tau \geq 1$, as shown before. Therefore, this cross term becomes sub-leading in $[\langle x_0 x_{0'} \rangle^2]$.

The cross term between the first and second terms vanishes as
\begin{eqnarray}
\left[ \left\langle \eta_{0,t} \sum_{\mu}\sum_{\tau \geq 1} \kappa_{0' \mu, t\ t-\tau} \phi_{\mu,t-\tau-1}\right\rangle^2\right]
&=& \left[  \left(\sum_{i,j,k}^N \sum_{\mu} \sum_{\tau \geq 1} J_{0i}J_{0'j}J_{k\mu} \left\langle\phi_{i,t}\phi'_{j,t} \phi_{\mu,t-\tau-1}\right\rangle \right)^2\right] \\
&=&  \left[  \left(\sum_{i,j,k}^N \sum_{\mu} \sum_{\tau \geq 1} J_{0i}J_{0'j}J_{k\mu} \left\langle\phi_{i,t}\phi'_{j,t} \right\rangle\left\langle \phi_{\mu,t}\right\rangle \right)^2\right] \nonumber \\
&=& 0, \nonumber
\end{eqnarray}
where the second equality holds because the activity of the reservoir neuron is independent of that of the auxiliary neurons.

Repeating these calculations, all cross terms are shown to scale as $O(1/N^2)$, except for the two $O(1/N)$ terms, $[\langle \eta_{0,t}\eta_{0',t} \rangle^2]$ and $[\langle (u_0 s_{t})(u_{0'}s_t)\rangle^2]= \tilde{\sigma}_s^4/N$.  Consequently, 
\begin{eqnarray}
[\langle x_0 x_{0'} \rangle^2] = \dfrac{g^4}{N}[\langle \phi_{i}^2 \rangle^2] + g^4 [\langle \phi_{i} \phi_{j} \rangle^2]_{(i\neq j)} + \frac{\tilde{\sigma}_s^4}{N} + O\left(\frac{1}{N^2}\right).\nonumber\\
\end{eqnarray}
Replacing $[\langle x_0 x_{0'} \rangle^2]$ with $[\langle x_i x_j \rangle^2]_{(i\neq j)}$, we obtain
\begin{eqnarray} \label{xx}
[\langle x_i x_j \rangle^2] = \dfrac{g^4}{N}[\langle \phi_{i}^2 \rangle^2] + g^4 [\langle \phi_{i} \phi_{j} \rangle^2]+ \frac{\tilde{\sigma}_s^4}{N} + O\left(\frac{1}{N^2}\right),\nonumber\\
\end{eqnarray}
for $i\neq j$. From the ansatz, $[\langle \phi_i \phi_j \rangle^2]_{(i\neq j)} \sim O(1/N)$, we see that $[\langle x_i x_j \rangle^2]_{(i\neq j)} \sim O(1/N)$. 

Consequently, similar to single-neuron statistics, the non-trivial perturbation induced by auxiliary neurons, i.e., the second and third terms in Eq.(\ref{AtoD for two cavity}), does not contribute to the leading order of neuronal correlations. In contrast, in the case of a continuous-time RNN, the perturbation induced by auxiliary neurons contributes to the leading order of neuronal correlations~\cite{Clark2022}. This discrepancy stems from the difference in the strength of the autocorrelation, $\langle x_i(t) x_i(t-\tau)\rangle_t$, which scales as $O(1)$ in the continuous-time model, but as $O(1/N^{\tau/2})$ in our discrete-time model. This property of negligible perturbation effects by auxiliary neurons in a discrete-time model greatly simplifies the analytical derivation of memory capacity (see Appendix~\ref{appendixB1}). Note that even in a discrete-time model, perturbation-induced effects contribute to the leading-order term of neuronal correlations if the RNN dynamics include a leak term, i.e. $x_i(t+1)=\gamma x_i(t)+\sum_j^N J_{ij}\phi_j(t)$ with $\gamma>0$, which yields an $O(1)$ autocorrelation.
 
To calculate $[\langle \phi_{i} \phi_{j} \rangle^2]$, we introduce Price's theorem~\cite{Helias}. For two Gaussian variables, $\begin{pmatrix} z_1 \\ z_2 \end{pmatrix} \sim \mathcal{N}\left(\begin{pmatrix} 0 \\ 0 \end{pmatrix}, \begin{pmatrix}c_1 & \tau \\ \tau & c_2 \\ \end{pmatrix} \right)$, and any smooth activation function $\phi$, we define $f_\phi(c_1,c_2,\tau) \equiv \langle \phi(z_1)\phi(z_2) \rangle$. Price's theorem then states:
\begin{eqnarray}
\partial_\tau f_\phi(c_1,c_2,\tau) = f_{\phi'}(c_1,c_2,\tau)     
\end{eqnarray}
Iteratively applying this theorem yields $\partial_\tau^n f_\phi(c_1,c_2,\tau) = f_{\phi^{(n)}}(c_1,c_2,\tau)$, which allows us to expand $\langle \phi(z_1)\phi(z_2)\rangle $ around $\tau=0$, resulting in
\begin{eqnarray} \label{Price's theorem}
\langle \phi(z_1)\phi(z_2) \rangle  &&= \sum_{n\geq 0} \frac{1}{n!} \tau^n   f_{\phi^{(n)}} (c_1, c_2, 0) 
= \sum_{n\geq 0} \frac{1}{n!}\langle z_1 z_2 \rangle^n \langle \phi^{(n)}(z_1) \rangle \langle \phi^{(n)}(z_2) \rangle.  
\end{eqnarray}

Applying Eq.(\ref{Price's theorem}) to $\langle \phi_i \phi_j \rangle$ yields
\begin{eqnarray}
\langle \phi_i \phi_j \rangle &&= \langle \phi'_{i,t}\rangle  \langle \phi'_{j,t} \rangle \langle x_i x_j \rangle + \frac{1}{3!} \langle \phi^{(3)}_{i,t}\rangle  \langle \phi^{(3)}_{j,t} \rangle \langle x_i x_j \rangle^3 + \frac{1}{5!} \langle \phi^{(5)}_{i,t}\rangle  \langle \phi^{(5)}_{j,t} \rangle \langle x_i x_j \rangle^5 + \cdots, 
\end{eqnarray}
where we used $\langle \phi^{(2n)}(x) \rangle = 0$ since $\phi^{(2n)}$ is an odd function. Therefore, we obtain
\begin{eqnarray} \label{phiphi}
[\langle \phi_i \phi_j \rangle^2] = [\langle \phi'_{i,t}\rangle^2]^2[\langle x_i x_j \rangle^2]+ O\left(\frac{1}{N^2}\right).
\end{eqnarray}
Substituting Eq.(\ref{phiphi}) into Eq.(\ref{xx}) yields
\begin{eqnarray}
[\langle x_i x_j \rangle^2] &&= \dfrac{g^4}{N}\langle \phi(x)^2 \rangle_*^2 + g^4 \langle \phi'(x)\rangle_*^4 [\langle x_i x_j \rangle^2] + \frac{\tilde{\sigma}_s^4}{N} + O\left(\frac{1}{N^2}\right),\nonumber \\
\end{eqnarray}
where $\langle f(x) \rangle_*$ denotes taking the average of $f(x)$ where $x\sim \mathcal{N}(0, [\langle x_i^2 \rangle])$. Therefore, we obtain 
\begin{eqnarray}\label{correlation}
[\langle x_i x_j \rangle^2] = \frac{1}{N} \frac{g^4\langle \phi(x)^2 \rangle_*^2 + \tilde{\sigma}_s^4}{1- g^4 \langle \phi'(x)\rangle_*^4} +O\left(\frac{1}{N^2}\right),
\end{eqnarray}
which is self-consistent with the ansatz, $[\langle \phi_i \phi_j\rangle^2]\sim O(1/N)$. 

\subsection{Proof of $0<g \langle \phi'(x) \rangle_* < 1$} 
\label{appendixA3}
We prove an inequality, $0<g \langle \phi'(x) \rangle_* < 1$ for any $g>0$, $\sigma_s^2 \geq 0$, $\sigma_n^2 \geq 0$, and an odd saturated sigmoid function satisfying $\phi'(0)=1$, $\phi'(x)>0$, $\phi''(x) \leq 0\ (x\geq 0)$ and $\phi(\pm \infty)=\pm 1$. We use the abbreviation, $K\equiv [\langle x_i^2 \rangle]$. It is evident that $g \langle \phi'(x) \rangle_* >0$ because $g>0$ and $\phi'(x) \geq 0$. We prove that $g\langle \phi'(x) \rangle_* < 1$ below. 

$\langle \phi'(x) \rangle_*$ is a decreasing function of $K$, because 
\begin{eqnarray}
\frac{d}{dK}\langle \phi'(x) \rangle_* &&= \frac{d}{dK}\int_{-\infty}^\infty \phi'(\sqrt{K}z) Dz 
= \frac{1}{\sqrt{K}}\int_0^\infty z \phi''(\sqrt{K}z) Dz \leq 0, 
\end{eqnarray}
where $Dz$ denotes a normal Gaussian measure. Since $K$ increases with both $\sigma_n^2$ and $\sigma_s^2$, it follows that $g \langle \phi'(x) \rangle_*$ decreases correspondingly. Therefore, it is enough to show that $g\langle \phi'(x) \rangle_* \leq 1$ only for the case of no input and noise, $\sigma_n^2=\sigma_s^2=0$. In this case, from Eq.(\ref{DMFeq}), $K$ is determined by solving
\begin{eqnarray} \label{DMFeq for no input}
K = g^2 \int \phi^2(\sqrt{K}z) Dz.    
\end{eqnarray}
For $g\leq1$, the only solution of this equation is $K=0$, resulting in $g\langle \phi'(x) \rangle_* = g\phi'(0)\leq 1$. For $g>1$, $K$ is uniquely determined as a function of $g$, so that $g$ can be assumed to be a function of $K$, 
\begin{eqnarray}
g^2 = \frac{K}{\int \phi^2(\sqrt{K}z) Dz}.
\end{eqnarray}
Therefore, it suffices to show the inequality,
\begin{eqnarray}
g^2 \left( \int \phi'(\sqrt{K}z) Dz \right)^2 = \frac{K \left( \int \phi'(\sqrt{K}z) Dz \right)^2}{\int \phi^2(\sqrt{K}z) Dz}
\leq 1, \nonumber\\
\end{eqnarray}
for any $K>0$. Employing integration by parts and Cauchy-Schwarz inequality, we can prove this inequality as
\begin{eqnarray}
K \left( \int \phi'(\sqrt{K}z) Dz \right)^2 
&&= \left( \int z \phi(\sqrt{K}z) Dz \right)^2
\leq \int z^2 Dz \int \phi^2(\sqrt{K}z) Dz 
= \int \phi^2(\sqrt{K}z) Dz.
\end{eqnarray}

\section{ANALYSIS ON MEMORY CAPACITY}
\label{appendixB}

\subsection{Analytical derivation of memory capacity} \label{appendixB1}

In this subsection, we derive the analytical solution of memory capacity. By simple calculation, $M_d$ defined in Eq.(\ref{Md}) can be expressed as
\begin{eqnarray} \label{appendix Md}
M_d = \frac{\bm{a}_d^\top C^{-1} \bm{a}_d}{\langle s(t)^2 \rangle},
\end{eqnarray}
where the elements of $\bm{a}\in \mathbb{R}^L$ and $C \in \mathbb{R}^{L\times L}$ are respectively $(\bm{a}_d)_i\equiv \langle s(t-d) x_i(t)\rangle$ and $C_{ij} \equiv \langle x_i(t) x_j(t)\rangle$ ($i,j$ indicate the indices of the readout neurons). Under the scaling assumption of model hyperparameters, Eq.(\ref{scaling}), the inverse of the covariance matrix $C$ can be expanded using Neumann series expansion:
\begin{eqnarray} \label{appendix neumann}
C^{-1} = \sum_{n=0}^\infty C_{\rm diag}^{-1}\left( - C_{\rm nondiag} C_{\rm diag}^{-1} \right)^n,
\end{eqnarray}
where $(C_{\rm diag})_{ij}\equiv \delta_{ij} \langle x_i(t)^2 \rangle$ and $(C_{\rm nondiag})_{ij}\equiv (1-\delta_{ij})\langle x_i(t) x_j(t) \rangle$. 

This series expansion is ensured to converge when the spectral norm is less than one, $\| C_{\rm nondiag} C_{\rm diag}^{-1}\| < 1$. However, calculating the exact value of the spectral norm is a challenging task. Employing the inequality $\| C_{\rm nondiag} C_{\rm diag}^{-1} \| \leq \| C_{\rm nondiag} C_{\rm diag}^{-1}\|_F$, where $\| \cdot \|_F$ denotes a Frobenius norm, we can obtain the sufficient condition for the convergence, $\| C_{\rm nondiag} C_{\rm diag}^{-1}\|_F <1$. The value of $\| C_{\rm nondiag} C_{\rm diag}^{-1} \|_F$ is calculated as
\begin{eqnarray}
\| C_{\rm nondiag} C_{\rm diag}^{-1}\|_F &&= \sum_{i,j(i\neq j)}^{\alpha\sqrt{N}} \frac{\langle x_i x_j \rangle^2}{\langle x_i^2 \rangle^2} 
= \alpha^2 N \left[ \frac{\langle x_i x_j \rangle^2_{(i\neq j)}}{\langle x_i^2 \rangle^2}\right] 
=\alpha^2 N  \frac{[\langle x_i x_j \rangle^2]_{(i\neq j)}}{[\langle x_i^2 \rangle]^2} 
= \frac{\alpha^2}{[\langle x_i^2 \rangle]^2} \frac{g^4\langle \phi(x)^2 \rangle_*^2 + \tilde{\sigma}_s^4}{1- g^4 \langle \phi'(x)\rangle_*^4},
\end{eqnarray}
where the second equality employs the self-averaging property, and the final equality utilizes Eq.(\ref{correlation}). In addition, the third equality follows from Eq.(\ref{mean of inverse}). Consequently, we obtain the sufficient condition,
\begin{eqnarray} \label{sufficient condition}
\alpha^2 < \frac{[\langle x_i^2 \rangle]^2}{\tilde{\sigma}_s^4 + ([\langle x_i^2 \rangle]-\sigma_n^2)^2}
\left(1-(g \langle \phi'(x) \rangle_*)^4 \right),
\end{eqnarray}
where we used the equation, $g^2 \langle \phi(x)^2 \rangle_* = [\langle x_i^2 \rangle]-\sigma_n^2$, derived from the dynamical mean-field equation, Eq(\ref{DMFeq appendix}).

Substituting the series expansion expression for $C^{-1}$ into Eq.(\ref{appendix Md}), we obtain 
\begin{eqnarray} \label{appendix Md2}
M_d =
\frac{\sqrt{N}}{\tilde{\sigma}_s^2} \sum_i^{\alpha \sqrt{N}}
\frac{\langle s_{t-d} x_{i,t} \rangle^2}{\langle x_i^2 \rangle}
-
\frac{\sqrt{N}}{\tilde{\sigma}_s^2} \sum_{\substack{i,j \\(i\neq j)}}^{\alpha \sqrt{N}}
\frac{\langle s_{t-d} x_{i,t} \rangle \langle x_i x_j \rangle \langle s_{t-d} x_{j,t} \rangle}{\langle x_i^2 \rangle \langle x_j^2 \rangle}
+
\frac{\sqrt{N}}{\tilde{\sigma}_s^2} \sum_{\substack{i,j,k\\(i\neq j, j\neq k)}}^{\alpha \sqrt{N}}
\frac{\langle s_{t-d} x_{i,t} \rangle \langle x_i x_j \rangle \langle x_j x_k \rangle  \langle s_{t-d} x_{k,t} \rangle}{\langle x_i^2 \rangle \langle x_j^2 \rangle \langle x_k^2 \rangle}
 - \cdots,\nonumber \\ 
\end{eqnarray}
where for simplicity we used the shorthand notation $x_{i,t}\equiv x_i(t)$, $s_t \equiv s(t)$, and $\langle x_i x_j \rangle \equiv \langle x_i(t) x_j(t) \rangle $.

In the large network size limit, we can assume self-averaging for $M_d$, that is, $\lim_{N\to \infty} M_d = [M_d] $, where the square bracket denotes the average over network realizations, known as "quenched average"~\cite{Helias}. In addition, as neurons exhibit statistically identical dynamics, the quenched averaging operation eliminates the dependence of each term in Eq.(\ref{Md}) on neuron indices $i,j,k,\cdots$. Consequently, Eq.(\ref{appendix Md2}) can be reduced to
\begin{eqnarray} \label{appendix Md3}
[M_d] = 
\frac{\alpha N}{\tilde{\sigma}_s^2}
\left[
\frac{\langle s_{t-d} x_{i,t} \rangle^2}{\langle x_i^2 \rangle}
\right] 
-
\frac{\alpha^2 N^{\frac{3}{2}}}{\tilde{\sigma}_s^2}
\left[
\frac{\langle s_{t-d} x_{i,t} \rangle \langle x_i x_j \rangle \langle s_{t-d} x_{j,t} \rangle}{\langle x_i^2 \rangle \langle x_j^2 \rangle}
\right]_{i\neq j} 
+
\frac{\alpha^3 N^2}{\tilde{\sigma}_s^2}
\left[
\frac{\langle s_{t-d} x_{i,t} \rangle \langle x_i x_j \rangle \langle x_j x_k \rangle  \langle s_{t-d} x_{k,t} \rangle}{\langle x_i^2 \rangle \langle x_j^2 \rangle \langle x_k^2 \rangle}
\right]_{\substack{i\neq j \\ j\neq k}} 
- \cdots.  \nonumber \\
\end{eqnarray}    

For the solution of memory capacity, it is enough to calculate each quenched averaging term in Eq.(\ref{appendix Md3}). We perform this calculation by the dynamical cavity method, similar to the approach described in appendix~\ref{appendixA}. As noted there, for a reservoir RNN with discrete time dynamics, perturbations induced by the auxiliary neurons do not contribute to the leading order term, which simplifies the calculation.

To begin with, we calculate the first term. Employing Eq.(\ref{mean of inverse}) and introducing an auxiliary neuron indexed by $0$, it is enough to calculate $[\langle s_{t-d} x_{0,t} \rangle^2]$ as:
\begin{eqnarray}
[\langle s_{t-d} x_{0,t} \rangle^2] &&= \left[\left\langle s_{t-d} \left( \sum_i^N J_{0i}\phi_{i,t-1}+u_0 s_t + \xi_{0,t}\right) \right\rangle^2\right]
= \sum_{i,j}^N \left[ J_{0i}J_{0j} \langle s_{t-d} \phi_{i,t-1}\rangle^2 \right] + [u_0^2]\langle s_{t-d}s_t \rangle^2 \nonumber \\
&&= \sum_{i,j}^N \left[ J_{0i}J_{0j}\right] \left[\langle s_{t-d} \phi_{i,t-1}\rangle^2 \right] + [u_0^2]\langle s_{t-d}s_t \rangle^2
= g^2 [\langle s_{t-(d-1)} \phi_{i,t} \rangle^2] + \delta_{d,0}\frac{\tilde{\sigma}_s^4}{N},
\end{eqnarray}
where we used the independence between $J_{0,i}, J_{0,j}$ and $\phi_{i,t}$ in the third equation. To proceed, we need to evaluate $[\langle s_{t-d} \phi(x_{i,t})\rangle^2]$. Since $(s_{t-d}, x_{i,t})$ are correlated Gaussian variables, we can express them using auxiliary variables $y, z$:
\begin{eqnarray}
    s_{t-d} &&= \sqrt{\sigma_s^2-\frac{\langle s_{t-d}x_{0,t}\rangle^2}{\langle x_0^2 \rangle}}\ y+ \frac{\langle s_{t-d}x_{0,t}\rangle}{\sqrt{\langle x_0^2 \rangle}}z \nonumber \\
    x_{0,t} &&= \sqrt{\langle x_0^2 \rangle}\ z.
\end{eqnarray}
Using these expressions, we can calculate $\langle s_{t-d} \phi(x_{i,t})\rangle$ as
\begin{eqnarray} \label{integration by parts}
    \langle s_{t-d} \phi(x_{i,t})\rangle &&= \int Dy\ Dz\ \left( \sqrt{\sigma_s^2-\frac{\langle s_{t-d}x_{0,t}\rangle^2}{\langle x_0^2 \rangle}}\ y+ \frac{\langle s_{t-d}x_{0,t}\rangle}{\sqrt{\langle x_0^2 \rangle}}z \right) \phi\left(\sqrt{\langle x_0^2 \rangle}\ z \right) \nonumber \\
    &&=  \frac{\langle s_{t-d}x_{0,t}\rangle}{\sqrt{\langle x_0^2 \rangle}} \int Dz\ z\  \phi\left(\sqrt{\langle x_0^2 \rangle}\ z \right) \nonumber \\
    &&= \langle s_{t-d}x_{0,t}\rangle \int Dz\ \phi'\left(\sqrt{\langle x_0^2 \rangle}\ z \right) \nonumber \\
    &&= \langle s_{t-d}x_{0,t}\rangle \langle \phi'(x_0) \rangle,
\end{eqnarray}
where $Dy$ and $Dz$ denotes standard Gaussian measures, and integration by parts is employed to derive the third line. Therefore, we obtain
\begin{eqnarray}
[\langle s_{t-d} x_{0,t} \rangle^2] = g^2[\langle \phi'_i \rangle^2][\langle s_{t-(d-1)}x_{0,t} \rangle^2] + \delta_{d,0}\frac{\tilde{\sigma}_s^4}{N}, \nonumber\\
\end{eqnarray}
where we restored $[\langle s_{t-(d-1)}x_{i,t} \rangle^2]$ to $[\langle s_{t-(d-1)}x_{0,t} \rangle^2]$. This equation is a recurrence formula for $[\langle s_{t-d} x_{0,t} \rangle^2]$, whose solution is
\begin{eqnarray} \label{sx}
[\langle s_{t-d} x_{0,t} \rangle^2] = \frac{1}{N}\tilde{\sigma}_s^4 \left(g\langle \phi'(x)\rangle_*\right)^{2d}.
\end{eqnarray}
Consequently, we obtain 
\begin{eqnarray}
{\rm 1st\ term\ of\ Eq.(\ref{appendix Md3}) }
= \frac{\alpha \tilde{\sigma}_s^2  \left(g\langle \phi'(x)\rangle_*\right)^{2d} }{[\langle x_i^2 \rangle]}.
\end{eqnarray}

Next, we move on to the calculation of the second term. We introduce the two auxiliary neurons indexed by $0$ and $0'$, obtaining for $d\geq 1$

\begin{eqnarray} \label{2nd term 1}
\left[ \langle s_{t-d} x_{0,t} \rangle \langle x_0 x_{0'} \rangle \langle s_{t-d} x_{0',t} \rangle\right]
&&=\left[ \sum_i^N J_{0 i} \langle s_{t-d} \phi_{i,t-1} \rangle 
\sum_{j,k}^N J_{0j}J_{0' k} \langle \phi_j \phi_k \rangle \sum_l^N J_{0' l} \langle \phi_{l,t-1} s_{t-d}  \rangle
\right]  \nonumber \\
&&=\sum_{i,j,k,l}^N \left[ J_{0 i} J_{0j} J_{0' k} J_{0' l}\langle s_{t-d} \phi_{i,t-1} \rangle \langle \phi_j \phi_k \rangle \langle \phi_{l,t-1} s_{t-d}  \rangle  \right] \nonumber \\
&&=\sum_{i,j,k,l}^N \left[ J_{0 i} J_{0j} J_{0' k} J_{0' l} \right] \left[ \langle s_{t-d} \phi_{i,t-1} \rangle \langle \phi_j \phi_k \rangle \langle \phi_{l,t-1} s_{t-d}  \rangle  \right] \nonumber \\
&&=\frac{g^4}{N^2} \sum_{i,j}^N \left[ \langle s_{t-d} \phi_{i,t-1} \rangle \langle \phi_i \phi_j \rangle \langle \phi_{j,t-1} s_{t-d} \rangle   \right] \nonumber \\
&&= \frac{g^4}{N}  [ \langle s_{t-(d-1)} \phi_{i,t} \rangle^2] [ \langle \phi_i^2 \rangle ] + g^4 \left[ \langle s_{t-(d-1)} \phi_{i,t} \rangle \langle \phi_i \phi_j \rangle \langle \phi_{j,t} s_{t-(d-1)} \rangle   \right]_{(i\neq j)}\nonumber \\
&&=  \frac{g^4}{N}  [ \langle  s_{t-(d-1)} \phi_{0,t} \rangle^2] [ \langle \phi_i^2 \rangle ] + g^4 [\langle \phi'(x_i) \rangle^2]^2 \left[ \langle s_{t-(d-1)} x_{0,t} \rangle \langle x_0 x_{0'} \rangle \langle x_{0',t} s_{t-(d-1)} \rangle  \right],\nonumber \\
\end{eqnarray}
where the last equality follows from Price's theorem. This equation is a recurrence formula for $\left[ \langle s_{t-d} x_{0,t} \rangle \langle x_0 x_{0'} \rangle \langle s_{t-d} x_{0',t} \rangle\right]$ with the initial condition ($d=0$),
\begin{eqnarray}
\left[ \langle s_{t} x_{0,t} \rangle \langle x_0 x_{0'} \rangle \langle s_{t} x_{0',t} \rangle\right] = 
\frac{\tilde{\sigma}_s^4}{N}[\langle x_0 x_{0'}\rangle u_0 u_{0'}]
=\frac{\tilde{\sigma}_s^4}{N}  \left[ u_0 u_{0'} \left(  \sum_{i,j}^N J_{0i}J_{0'j} \langle \phi_i \phi_j \rangle + u_0 u_{0'} \langle s_t^2 \rangle 
\right) \right]
= \tilde{\sigma}_s^6 N^{-3/2}.
\end{eqnarray}
Thus, the second term of Eq.(\ref{2nd term 1}) for $d=1$ is $O(N^{-3/2})$. On the other hand, from Eq.(\ref{sx}), the first term of Eq.(\ref{2nd term 1}) for $d=1$ scales as $O(N^{-2})$. As a result, this term becomes subleading and negligible. Consequently, solving the recurrence formula yields
\begin{eqnarray}
\left[ \langle s_{t-d} x_{0,t} \rangle \langle x_0 x_{0'} \rangle \langle s_{t-d} x_{0',t} \rangle\right]
= \frac{\tilde{\sigma}_s^6  \left(g\langle \phi'(x)\rangle_*\right)^{4d} }{N^{3/2}} + {\rm subleading\ terms}
\end{eqnarray}
and thus
\begin{eqnarray}
{\rm 2nd\ term\ of\ Eq.(\ref{appendix Md3}) }
= - \frac{\alpha^2 \tilde{\sigma}_s^4  \left(g\langle \phi'(x)\rangle_*\right)^{4d} }{[\langle x_i^2 \rangle]^2}+ {\rm subleading\ terms}
\end{eqnarray}

Subsequently, we proceed to calculate the third term in Eq.(\ref{appendix Md3}). It is enough to consider the case for $i\neq k$ because the terms for $i=k$ have the same order as those for $i\neq k$, but the number of $i=k$ terms, $O(N^2)$, is much smaller than those of $i\neq k$ terms, $O(N^3)$, making their overall contribution subleading. We introduce three auxiliary neurons, labeled $0$, $0'$, and $0''$, obtaining for $d\geq 1$
\begin{eqnarray}\label{sx 3rd term}
\left[\langle s_{t-d} x_{0,t} \rangle \langle x_0 x_{0'} \rangle \langle x_{0'} x_{0''} \rangle \langle x_{0'',t} s_{t-d} \rangle \right]
&&= \left[ \sum_i^N J_{0i} \langle s_{t-d}\phi_{i,t-1} \rangle \sum_{j,k}^N J_{0j}J_{0' k} \langle \phi_j \phi_k \rangle \sum_{l,m}^N J_{0'l}J_{0''m } \langle \phi_l \phi_m \rangle \sum_n^N J_{0''n} \langle \phi_{n,t-1} s_{t-d} \rangle  \right] \nonumber \\
&&= \frac{g^6}{N^3} \sum_{i,j,k}^N \left[ \langle s_{t-d} \phi_{i,t-1} \rangle \langle \phi_i \phi_j \rangle \langle \phi_j \phi_k \rangle \langle \phi_{k,t-1} s_{t-d} \rangle \right] \nonumber \\
&&=
\frac{g^6}{N^2}[\langle \phi_i^2 \rangle]^2 [\langle s_{t-(d-1)} \phi_{i,t} \rangle^2] + \frac{2 g^6}{N} [\langle \phi_i^2 \rangle] \left[ \langle s_{t-(d-1)} \phi_{0,t} \rangle \langle \phi_0 \phi_{0'} \rangle \langle \phi_{0',t} s_{t-(d-1)} \rangle \right] \nonumber \\
&&\hspace{60pt}+ g^6  [\langle \phi'_i \rangle^2]^3 [\langle s_{t-(d-1)} x_{0,t} \rangle \langle \phi_0 \phi_{0'} \rangle \langle x_{0'} x_{0''} \rangle \langle x_{0'',t} s_{t-(d-1)} \rangle ],
\end{eqnarray}
where we used Price's theorem in the last equality. This equation is a recurrence formula for $\left[\langle s_{t-d} x_{0,t} \rangle \langle x_0 x_{0'} \rangle \langle x_{0'} x_{0''} \rangle \langle x_{0'',t} s_{t-d} \rangle \right]$ with initial condition ($d=0$),
\begin{eqnarray}
\left[\langle s_{t} x_{0,t} \rangle \langle x_0 x_{0'} \rangle \langle x_{0'} x_{0''} \rangle \langle x_{0'',t} s_{t} \rangle \right]
&&= \frac{\tilde{\sigma}^4}{N} [\langle x_0 x_{0'} \rangle \langle x_{0'} x_{0''} \rangle u_0 u_{0''} ] \nonumber \\
&&= \frac{\tilde{\sigma}^4}{N}  \left[ u_0 u_{0''} \left(  \sum_{i,j}^N J_{0i}J_{0'j} \langle \phi_i \phi_j \rangle + u_0 u_{0'} \langle (s_t)^2 \rangle  \right)  \left(  \sum_{i,j}^N J_{0'i}J_{0''j} \langle \phi_i \phi_j \rangle + u_{0'} u_{0''} \langle (s_t)^2 \rangle  \right) 
\right] \nonumber \\
&&= \tilde{\sigma}^8 N^{-2}.
\end{eqnarray}

Therefore, the third term in Eq.(\ref{sx 3rd term}) for $d=1$ is $O(1/N^2)$, while the first and second term scale $O(1/N^3)$ and $O(1/N^{5/2})$, respectively. As a result, only third term contributes to the leading order. Solving the recurrence formula, Eq.(\ref{sx 3rd term}), to leading order, we obtain
\begin{eqnarray}
\left[\langle s_{t} x_{0,t} \rangle \langle x_0 x_{0'} \rangle \langle x_{0'} x_{0''} \rangle \langle x_{0'',t} s_{t} \rangle \right] 
=  \frac{\tilde{\sigma}_s^8  \left(g\langle \phi'(x)\rangle_*\right)^{6d} }{N^2} + {\rm subleading\ terms}
\end{eqnarray}
and thus
\begin{eqnarray}
{\rm 3rd\ term\ of\ Eq.(\ref{appendix Md3}) }
= \frac{\alpha^3 \tilde{\sigma}_s^6  \left(g\langle \phi'(x)\rangle_*\right)^{6d} }{[\langle x_i^2 \rangle]^3}+ {\rm subleading\ terms}
\end{eqnarray}

The analogous calculation can be applied to the remaining terms, and we finally obtain the analytical solution for $M_d$ and memory capacity:
\begin{eqnarray}\label{mc appendix}
M_d = \sum_{n=0}^\infty
(-1)^{n}
\left\{
\frac{\alpha \tilde{\sigma}_s^2}{[\langle x_i^2\rangle]}  
\left(
g\langle \phi'(x) \rangle_{*}
\right)^{2d}
\right\}^{n+1}, 
\hspace{20pt}
MC = \sum_{d=0}^\infty \sum_{n=0}^\infty
(-1)^{n}
\left\{
\frac{\alpha \tilde{\sigma}_s^2}{[\langle x_i^2\rangle]}  
\left(
g\langle \phi'(x) \rangle_{*}
\right)^{2d}
\right\}^{n+1}. \nonumber\\
\end{eqnarray}

\subsection{Generalization of scaling assumptions} \label{appendixB2}

In this subsection, our aim is to identify broader scaling assumptions -- beyond those specified in Eqs.(\ref{scaling}) -- under which our theoretical framework for analytically deriving memory capacity remains applicable. To this end, we consider the following generalized scaling relations:
\begin{eqnarray}
    \label{Eq:generalized scaling}
    L=\alpha N^{c_1},\ \ \sigma_s^2 = \frac{\tilde{\sigma}_s^2}{N^{c_2}},\ \ 
\sigma_n^2 \sim O(1),
\end{eqnarray}
where $c_1, c_2 \geq 0$. 

First, we require that the memory capacity remains non-vanishing in the limit  $N\to\infty$. The first-order term in the Neumann series expansion of $[M_d]$ is given by
\begin{eqnarray}
    \frac{1}{\sigma_s^2}\sum_{i}^{\alpha N^{c_1}} \left[ \frac{\langle s_{t-d}x_{i,t} \rangle^2}{\langle x_i^2 \rangle}\right] \sim O(N^{c_1-c_2}).
\end{eqnarray}
To ensure that the memory capacity does not vanish as $N\to \infty$, it is necessary that $c_1\geq c_2$. 

Second, from the sufficient condition for the convergence of the Neumann series expansion, we obtain an additional constraint on $c_1$ and $c_2$. Specifically, the Frobenius norm of the matrix $C_{\rm nondiag} C_{\rm diag}^{-1}$ must satisfy
\begin{eqnarray}
    \label{Eq:order condition of frobenius norm}
    \| C_{\rm nondiag} C_{\rm diag}^{-1} \|_F =  \alpha^2 N^{2c_1}  \frac{[\langle x_i x_j \rangle^2]_{(i\neq j)}}{[\langle x_i^2 \rangle]^2} \sim O(1).
\end{eqnarray}
The order of the neuronal correlations $[\langle x_i x_j \rangle^2]_{(i\neq j)}$ can be obtained using the same calculation as described in appendix~{\ref{appendixA2}}, yielding $[\langle x_i x_j \rangle^2]_{(i\neq j)} \sim O(1/N^{\min(1, 2c_2)})$. Substituting this result into Eq.(\ref{Eq:order condition of frobenius norm}), we obtain
\begin{eqnarray}
    \| C_{\rm nondiag} C_{\rm diag}^{-1} \|_F \sim  O(N^{2c_1-\min({1, 2c_2})}),
\end{eqnarray}
from which it follows that $2c_1 \leq \min(1,2c_2)$ must be satisfied.

To summarize, the scaling exponents, $c_1$ and $d_2$, must satisfy the following two conditions:
\begin{eqnarray}
    c_1 &&\geq c_2 \\
    2c_1 &&\leq \min(1,2c_2), \nonumber
\end{eqnarray}
Combining these inequalities, we conclude that the allowed values are $0 \leq c_1=c_2 \leq \frac{1}{2}$.

\subsection{Analytical solution of the decay rate, $\lim_{N\to\infty} r(L)$} \label{appendixB3}

Noting that by applying Eq.(\ref{mc appendix}), the denominator in the definition of the decay rate, Eq.(\ref{decay rate}), can be calculated as
\begin{eqnarray}
\lim_{N\to\infty} L\times MC(1) 
&&\overset{{\rm Eq}.(\ref{mc appendix})}{=}
\lim_{N\to\infty} \alpha\sqrt{N}\times 
 \sum_{d=0}^\infty \sum_{n=0}^\infty
(-1)^{n} \left\{ \frac{1}{\sqrt{N}}
\frac{\tilde{\sigma}_s^2}{[\langle x_i^2\rangle]}  
\left(
g\langle \phi'(x) \rangle_{x\sim \mathcal{N}(0,[\langle x_i^2\rangle])}
\right)^{2d}
\right\}^{n+1} \nonumber \\
&&= \frac{\alpha \tilde{\sigma}_s^2}{[\langle x_i^2\rangle] \left(1-g^2 \langle \phi'(x) \rangle_*^2 \right)},
\end{eqnarray}
the decay rate of memory capacity can be analytically derived as
\begin{eqnarray}
\lim_{N\to\infty}r(L=\alpha\sqrt{N}) = \frac{[\langle x_i^2\rangle] \left(1-g^2 \langle \phi'(x) \rangle_*^2 \right)}{\alpha \tilde{\sigma}_s^2}
\sum_{d=0}^\infty \sum_{n=0}^\infty
(-1)^{n}
\left\{
\frac{\alpha \tilde{\sigma}_s^2}{[\langle x_i^2\rangle]}  
\left(
g\langle \phi'(x) \rangle_{*}
\right)^{2d}
\right\}^{n+1}.
\end{eqnarray}
The two infinite summations, $\sum_{d=0}^\infty$ and $\sum_{n=1}^\infty$, can be swapped under the condition, Eq.(\ref{sufficient condition}), because the infinite series is absolutely convergent:  
\begin{eqnarray}
\sum_{d=0}^\infty \sum_{n=0}^\infty
\left\{
\frac{\alpha \tilde{\sigma}_s^2}{[\langle x_i^2\rangle]}  
\left(
g\langle \phi'(x) \rangle_{*}
\right)^{2d}
\right\}^{n+1} 
= \sum_{d=0}^\infty \frac{\frac{\alpha \tilde{\sigma}_s^2}{[\langle x_i^2\rangle]}  \left(g\langle \phi'(x) \rangle_{*} \right)^{2d}}{1-\frac{\alpha \tilde{\sigma}_s^2}{[\langle x_i^2\rangle]}  \left(g\langle \phi'(x) \rangle_{*} \right)^{2d}} 
< \sum_{d=0}^\infty \frac{\alpha \tilde{\sigma}_s^2}{[\langle x_i^2\rangle]}  \left(g\langle \phi'(x) \rangle_{*} \right)^{2d} 
< \frac{\alpha \tilde{\sigma}_s^2}{[\langle x_i^2\rangle]} \frac{1}{1-(g\langle \phi'(x)\rangle_*)^2} < \infty. \nonumber\\
\end{eqnarray}
Here, in the first inequality, we apply the result $0<g\langle \phi'(x) \rangle_* < 1$ proven in appendix~\ref{appendixA3}, and  $\frac{\alpha \tilde{\sigma}_s^2}{[\langle x_i^2\rangle]}<1$ as shown below:
\begin{eqnarray}
\frac{\alpha \tilde{\sigma}_s^2}{[\langle x_i^2\rangle]}
< \frac{\tilde{\sigma}_s^2}{[\langle x_i^2\rangle]} \sqrt{\frac{[\langle x_i^2 \rangle]^2}{\tilde{\sigma}_s^4 + ([\langle x_i^2 \rangle]-\sigma_n^2)^2}
\left(1-(g \langle \phi'(x) \rangle_*)^4 \right)} 
< \frac{\tilde{\sigma}_s^2}{[\langle x_i^2\rangle]} \sqrt{\frac{[\langle x_i^2 \rangle]^2}{\tilde{\sigma}_s^4}} = 1,
\end{eqnarray}
where we used Eq.(\ref{sufficient condition}) in the first inequality.

Consequently, by interchanging the two infinite summations, the decay rate can be represented as
\begin{eqnarray} \label{appendix analytical r}
\lim_{N\to\infty}r(L=\alpha\sqrt{N}) = 1 - \sum_{n=1}^\infty (-1)^{n-1}\left(\frac{\tilde{\sigma}_s^2}{[\langle x_i^2 \rangle]} \alpha \right)^{n} \frac{1- (g\langle \phi'(x) \rangle_*)^2}{1- (g\langle \phi'(x) \rangle_*)^{2n+2}},
\end{eqnarray}
where the second term arises due to the neuronal correlations.

\subsection{Decay rate for a linear RNN} \label{appendixB4}

We mention the decay rate of memory capacity for the reservoir RNN with a linear activation function, $\phi(x)=x$. In this case, the decay rate can be derived as 
\begin{eqnarray}
\lim_{N\to\infty}r(L=\alpha\sqrt{N}) = 1 - \sum_{n=1}^\infty (-1)^{n-1}\left( \frac{\tilde{\sigma}_s^2}{\sigma_n^2} \alpha \right)^{n} \frac{(1-g^2)^{n+1}}{1-g^{2n+2}},
\end{eqnarray}
where the value of $g$ must be less than one to ensure the stability of the reservoir RNN. Analogous to non-linear RNNs, the memory capacity of the linear RNN with neuronal noise exhibits sublinear scaling with respect to $L$, as illustrated in Fig.\ref{fig:decaying rate for linear RNN}. In contrast, for a linear RNN in the absence of noise, the memory capacity is known to be exactly $L$~\cite{Jaeger2002}, resulting in $r(L)=1$. It is intriguing that the presence or absence of noise greatly affects the scaling behavior of memory capacity with respect to $L$. 

Here, we mention several points. First, the sublinear scaling of memory capacity for a linear RNN subject to neuronal noise is not accompanied by an increase in nonlinear computational capabilities, unlike in the case of a nonlinear RNN (Fig.\ref{fig:ipc}(a)(b)). This is simply because a linear RNN lacks the ability to perform nonlinear computations. Consequently, in a linear RNN, the sublinear scaling of memory capacity solely reflects a decline in overall computational performance caused by neuronal noise. Second, our theoretical analysis based on the Neumann series expansion is not applicable to the noise-free linear RNN. In this setting, the diagonal entries of covariance matrix $C$ are comparable to its non-diagonal entries, both scaling as $O(1/\sqrt{N})$, which prevents us from employing the Neumann series expansion.

\begin{figure}[htbp]
    \centering
    \includegraphics[width=0.7\linewidth]{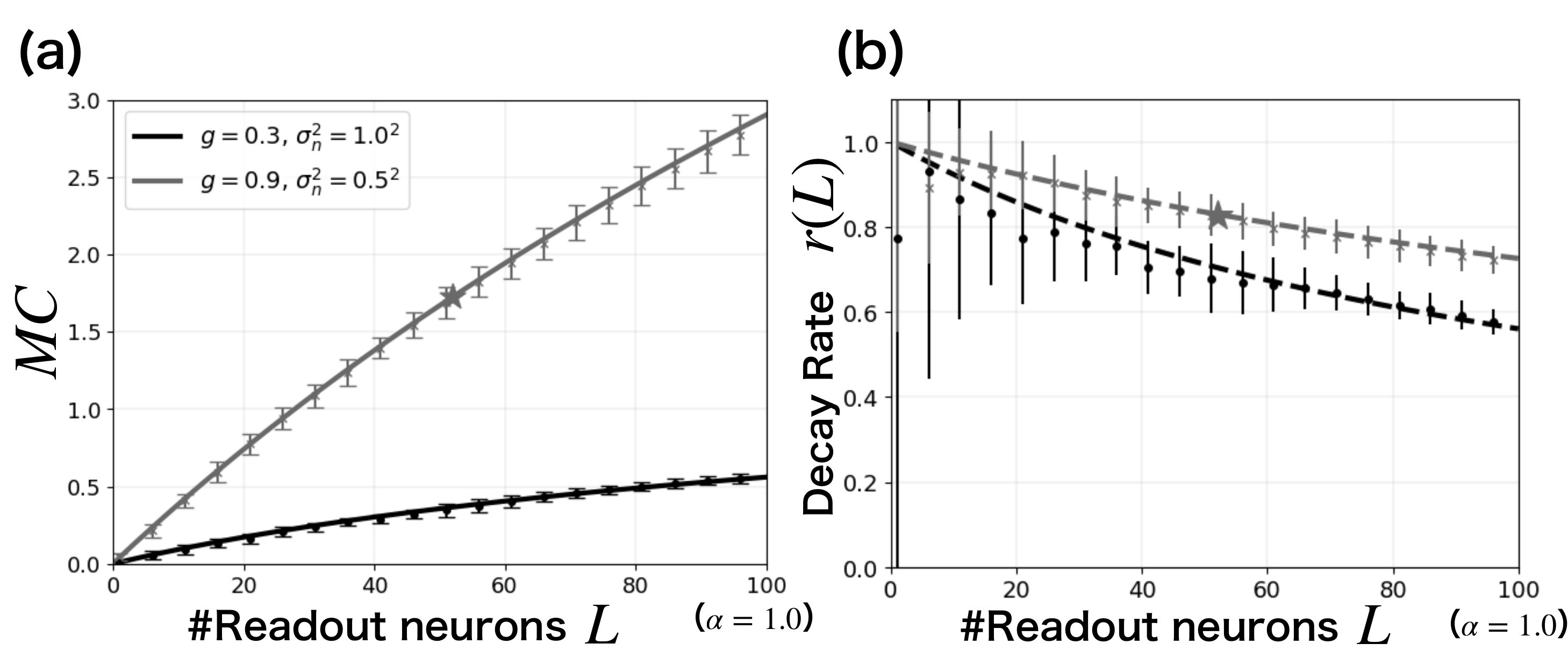}
    \caption{ (a) Memory capacity and (b) growth rate of memory capacity for linear RNNs. The figure's structure and content are the same as Fig.\ref{fig:analytical MC}, with analytical and numerical results presented in the same format. The input intensity is $\sigma_s^2 = 1.0^2 / \sqrt{N}$ ($\tilde{\sigma}_s^2 = 1.0^2$). In simulations, the network size is $N=10000$, and simulation time length is $T=10^4$. 
    }
    \label{fig:decaying rate for linear RNN}
\end{figure}

\section{INFORMATION PROCESSING CAPACITY THEORY}
\label{appendixC}

\subsection{Overview of IPC theory} \label{appendixC1}

We provide a concise overview of the IPC, including its definition and key properties. We consider a reservoir receiving input signals $\{ s(t) \}_t$. The general task is formulated as optimizing output weights such that the reservoir's output $z(t)$ approximates a given function of the inputs, $f[\cdots, s(t-1), s(t)]$. For input signals drawn i.i.d from a standard Gaussian distribution, the orthonormal basis functions spanning the Hilbert space containing $f$ are expressed as a formally infinite product of normalized Hermite polynomials:
\begin{eqnarray} \label{definition y_d}
y_{\bm{d}} = \prod_{i\geq 0}\mathcal{H}_{d_i}\left( s(t-i) \right),
\end{eqnarray}
where $\mathcal{H}_{d_i}(\cdot)$ ($d_i\geq 0$) denotes the normalized Hermite polynomial of degree $d_i$, and vector-form index $\bm{d}\equiv (d_i)_{i\geq 0}$ has only a finite number of non-zero elements. Since $\mathcal{H}_0= 1$, the product in Eq.(\ref{definition y_d}) is effectively finite. The summation of $d_i$ is equivalent to the degree of the polynomial $y_{\bm{d}}$, defined as $deg(y_{\bm{d}}) \equiv \sum_{i\geq 0} d_i$. 

The {\it{Capacity}} for the reservoir to reconstruct the function of inputs, $f$, is defined as 
\begin{eqnarray} \label{capacity}
C[f] \equiv 1-\frac{{\rm{min}}_{\bm{w}}\langle (z(t)-f(t))^2 \rangle_t}{\langle f(t)^2 \rangle_t},
\end{eqnarray}
where $\langle \cdot \rangle_t$ denotes the time average. It is proven that $0\leq C[f] \leq 1$ holds true for any function $f$. The {\it{total IPC}} is subsequently defined as the summation of capacities across all basis polynomials: 
\begin{eqnarray}
IPC_{\rm{total}} \equiv \sum_{\bm{d}} C[y_{\bm{d}}].
\end{eqnarray}
Notably, it has been established that $IPC_{\rm total}=L$ holds, provided that the reservoir's activity is full-rank and the reservoir satisfies the {\it fading memory property}~\cite{Dambre2012}, whereby the reservoir's state is uniquely determined solely by input signals, regardless of its initial state. 

The total IPC can be decomposed based on the degree of the basis polynomials. We define the {\it{IPC for degree $D$}} as 
\begin{eqnarray}
IPC_{D} \equiv \sum_{\substack{\bm{d} \\ {\rm{s.t.}}\ deg(y_{\bm{d}})=D}} C[y_{\bm{d}}].
\end{eqnarray}
Crucially, $IPC_1$ is identical to memory capacity, as $C[y_{\bm{d}}]$ for $deg(y_{\bm{d}})=1$ corresponds to Eq.(\ref{Md}). In contrast, $IPC_D$ for $D \geq 2$ represents a non-linear computational ability of the reservoir. 

When calculating the capacities by numerical simulation, we must take care not to overestimate them, because for a finite simulation time $T$, they are subject to a systematic positive error. Following Dambre et al.~\cite{Dambre2012}, we determine the threshold of the capacities, $\epsilon$, as
\begin{eqnarray} \label{cutoff}
\epsilon \equiv \frac{2\theta}{T},\ \ \theta \equiv \arg_\theta \left\{\mathbb{P}[\chi^2(L) \geq \theta] = p \right\},
\end{eqnarray}
where $\chi^2(L)$ denotes a random variable that follows a Chi-squared distribution with $L$ degrees of freedom. The value of $p$ represents the probability that a truly zero capacity is incorrectly assessed as non-zero. We consider a capacity to be zero if it falls below this threshold $\epsilon$. Consistent with Dambre et al., we set $p = 10^{-4}$ for all our numerical simulations of memory capacity ($=IPC_1$) and IPC, although the choice of $p$ has a negligible impact on our results. 

Note that calculating the IPC values of our reservoir RNN model requires a minor correction to the time evolution equation, Eq.\ref{model}, as 
\begin{eqnarray}
x_i(t) = \sum_{j=1}^N J_{ij} \phi(x_i(t-1)) +  u_i \sigma_s s(t) + \xi_i(t),
\ \ s(t)\sim \mathcal{N}(0,1),
\end{eqnarray}
where the standard deviation of the input signals, $\sigma_s$, is now incorporated as a scaling factor of the standard Gaussian noise inputs. This correction is necessary because IPC is defined for a reservoir receiving standard Gaussian noise inputs. Importantly, this modification does not alter the statistical behavior or computational ability of the original RNN model.

\subsection{Linear scaling of $IPC_D$ with $L$ for vanishing neuronal correlations}
\label{appendixC2}
In the main text, we claim that each $IPC_D$ for $D\geq 1$ exhibits linear scaling with regard to $L$ if we ignore neuronal correlations. The following provides a concise proof of this claim. First, similarly to Eq.(\ref{Md2}), the capacity for a basis function $y_{\bm d}$ can be expressed as 
\begin{eqnarray}
C_T[y_{\bm d}] = \frac{\bm{a}^\top C^{-1} \bm{a}}{\langle y_{\bm d}(t)^2 \rangle},
\end{eqnarray}
where the elements of $\bm{a} \in \mathbb{R}^L$ are $a_i = \langle y_{\bm d}(t) x_i(t) \rangle $ and the matrix $C\in \mathbb{R}^{L\times L}$ is the covariance matrix of the readout neurons. Assuming neuronal correlations vanish, the capacity for any $y_{\bm{d}}$ is proportional to $L$ since
\begin{eqnarray}
C_T[y_{\bm{d}}] &&\approx \left[ C_T[y_{\bm{d}}] \right]
\approx \left[ \sum_i^L \frac{\langle y_{\bm{d}}(t) x_i(t)^2 \rangle}{\langle x_i(t)^2 \rangle \langle y_{\bm d}(t)^2 \rangle } \right ] 
= L \left[\frac{\langle y_{\bm{d}}(t) x_i(t) \rangle^2}{\langle x_i(t)^2 \rangle \langle y_{\bm d}(t)^2 \rangle} \right]
\propto L,
\end{eqnarray}
where we assume the self-averaging of the capacity in the first approximation equation, and in the second approximation equation, we ignore the correlation $\langle x_i x_j \rangle$.

\end{widetext}

\end{document}